\documentclass[11pt, secnumarabic,amssymb, nobibnotes, aps, prd]{revtex4-2}
	
\usepackage[dvips]{graphicx,color}
\usepackage{times}
\usepackage{filecontents}
\usepackage{mathtools}
\usepackage{subcaption}
\usepackage{graphicx}
\usepackage{amsmath}
\usepackage{multirow}
\usepackage{xcolor}
\usepackage{float}
\usepackage{nccmath}
\usepackage{amsmath}

\usepackage[%
  colorlinks=true,
  urlcolor=blue,
  linkcolor=red,
  citecolor=blue
]{hyperref}

\begin{document}
\title{Thermodynamic topology of 4D Euler-Heisenberg-AdS black hole in different ensembles}

\author{Naba Jyoti Gogoi$^1$  }\email{gogoin799@gmail.com }
\author{Prabwal Phukon$^{1,2}$} \email{prabwal@dibru.ac.in}

\affiliation{$1.$ Department of Physics, Dibrugarh University,
Dibrugarh 786004, Assam, India \\
$2.$ Theoretical physics division, Centre for atmospheric studies, Dibrugarh University}

\begin{abstract}
We study the thermodynamic topology of 4D Euler-Heisenberg-AdS (EHAdS) black hole and higher-order QED corrected Euler-Heisenberg-AdS black hole in different ensembles using generalized off-shell free energy. In this approach, black holes are viewed as defects in the thermodynamic space. We work in two ensembles: canonical ensemble in which the charge is kept fixed and grand canonical ensemble in which the conjugate potential $\phi_e$ is kept fixed. In each case, the local and global topology of the thermodynamic space is investigated via the computation of winding numbers at the defects. For 4D Euler-Heisenberg-AdS black hole in canonical ensemble, the topological class is found to be different depending on the Euler-Heisenberg (EH) parameter $a$. The topological numbers for $a<0$ and $a>0$ cases are found to be $W=+1$ and $W=0$ respectively. The topological number is found to be independent of the variation in pressure $P$ and charge $Q$ of the black hole. With the introduction of higher order QED correction, the difference in the topological class of 4D EHAdS black hole with the sign of $a$ is observed to go away.The topological number in this case is found to be $W=+1$ irrespective of the values of $a$, $P$ and $Q$. In the grand canonical ensemble, the topological number for both EHAdS and higher order QED corrected EHAdS black hole is found to be $W=0$, independent of the values of $P$, $\phi_e$ and $a$. Therefore, we infer that the topological class of both 4D EHAdS black hole and higher order QED corrected EHAdS black hole is ensemble dependent. Moreover, in the canonical ensemble, higher order QED correction alters the topological class of the black hole for positive values of EH parameter $a$. In the grand canonical ensemble, the higher order corrections do not change the thermodynamic topology of the black hole.

\end{abstract}

\pacs{04.30.Tv, 04.50.Kd}
\keywords{Black Hole Thermodynamics, Topology, Topological defect, R-charged black holes}

\maketitle

\author{Naba Jyoti Gogoi$^1$  }\email{gogoin799@gmail.com }
\author{Prabwal Phukon$^{1,2}$} \email{prabwal@dibru.ac.in}

\affiliation{$1.$ Department of Physics, Dibrugarh University,
Dibrugarh 786004, Assam, India \\
$2.$Theoretical physics division, Centre for atmospheric studies, Dibrugarh University}

\section{Introduction}
Since the realization of black holes as thermodynamic systems, it has been a very active area of research in physics producing many foundational and interesting results \cite{Hawking, Bekenstein73, Hawking:1974sw, Bekenstein:1972tm, Bekenstein:1973ur, Bardeen:1973gs, Phy, bekenstein1980black, Wald:1999vt, Carlip:2014pma, Wall:2018ydq, Candelas:1977zz, Chamblin:1999hg, Hawking:1982dh, Chamblin:1999tk}. Of late, black hole thermodynamics has been rigorously studied in the extended thermodynamic space where the cosmological constant $\Lambda$ is incorporated in the Smarr formula as pressure $P$ \cite{Kastor:2009wy,Dolan:2012jh, Gunasekaran:2012dq,Chen:2016gzz}. The thermodynamics and phase transition of various anti-de Sitter (AdS) black holes have been studied in this space \cite{Kubiznak:2012wp,Altamirano:2013ane,Altamirano:2013uqa,Wei:2014hba,Frassino:2014pha,Cai:2013qga,Xu:2014tja,Dolan:2014vba,Hennigar:2015esa,Hennigar:2015wxa,Hennigar:2016xwd,Zou:2016sab,Gogoi:2021syo,Gogoi:2023qni}.

Another important contribution to this area of research is the topological approach of black hole thermodynamics. The idea was proposed in \cite{Wei:2021vdx} which is a simple and straightforward method adopted from Duan's topological $\phi$-mapping theory\cite{Duan}. Two formerly unknown critical points: novel and conventional, are proposed and are endowed with topological charges $+1$ and $-1$ respectively. The conventional critical point is related to the first-order phase transition of a black hole system. Following this, another important concept is proposed in \cite{Wei:2022dzw} where, using the generalized off-shell free energy, the black hole systems are viewed as thermodynamic defects in the parameter space. In this approach, the generalized free energy expression is used which is given as
\begin{equation}
\mathcal{F}=E-\frac{S}{\tau},
\end{equation}
where, $E$ and $S$ are the energy and entropy of the black hole. Also, $\tau$ is the inverse of temperature. The vector function is calculated as $\phi=(\partial_{r_+}\mathcal{F},-\cot\Theta\csc\Theta)$. The $\Theta$ parameter is added for convenience. The condition $\Theta=\pi/2$ gives the zero points of $\phi$. Around each zero point contour $C$ parametrized by $\vartheta \in (0,2\pi)$ is drawn which is defined as \cite{Wei:2021vdx}:
\begin{equation}	
\label{Contour}
	\begin{cases}
		&r_+=a\cos\vartheta+r_0, \\
		&\theta=b\sin\vartheta+\frac{\pi}{2}.
	\end{cases}
\end{equation}
The winding numbers are calculated by $w=\frac{1}{2\pi}\Omega(2\pi)$ where the deflection of vector field $n$ along the contour $C$ is 
\begin{equation}
\label{Deflection Angle}
\Omega(\vartheta)=\int_0^\vartheta \epsilon_{ab}n^a\partial_\vartheta n^b  d\vartheta.
\end{equation}
The winding numbers reveal the local topological behaviour at the defects and the topological number $W$, which is the sum of the winding numbers, provides information about the global topological behaviour. The study of thermodynamic topology by now has been extended to a number of black hole systems \cite{Wu:2022whe,Liu:2022aqt,Fan:2022bsq,Gogoi:2023qku,Gogoi:2023xzy,Ye:2023gmk,Zhang:2023uay,Du:2023wwg,Sharqui:2023mbx,Du:2023nkr,Wu:2023sue,Wu:2023xpq,Ali:2023zww,Sadeghi:2023dsg,Saleem:2023oue,Shahzad:2023cis,Chen:2023elp,Bai:2022klw,Yerra:2022alz,Hazarika:2023iwp,Mehmood:2023psa,Tong:2023lyc,Wang:2023qxw,Sadeghi:2023tuj,Wu:2023fcw,Wu:2023meo}.

Recently, an alternative method of calculating the winding numbers is proposed by Fang, Jiang and Zhang which uses the concept of residue theorem \cite{Fang:2022rsb}. This method suggest a complex function $\mathcal{R}(z)$ which is defined as 
\begin{equation}
\label{Eq:GeneralComplexFunction}
\mathcal{R}(z)\equiv \frac{1}{\tau-\mathcal{G}(z)}.
\end{equation}
Here, $\tau=\mathcal{G}(r_+)$ is calculated from the generalized free energy using the condition $\partial_{r_+}\mathcal{F}=0$. Then, the event horizon radius $r_+$ is replaced by a complex variable $z$. With such consideration, the defect points can be represented by the isolated singular points $z_i$ of $\mathcal{R}(z)$ on the complex plane $\mathbb{C}$ defined by complex variable $z=x+iy$ with $x,y \in \mathbb{R}$. Contour $C_i$ can be drawn enclosing each real singular point $z_i=z_1,z_2,...,z_m$. The complex function $\mathcal{R}(z)$ is then analytic on and within the contour except for these singular points. Using the Residue theorem, integral of $\mathcal{R}(z)$ is calculated along each contour $C_i$
\begin{equation}
\oint_C\frac{\mathcal{R}(z)}{2\pi i}dz=\sum_{k=1}^m \text{Res}[\mathcal{R}(z_k)].
\end{equation}
Then, the winding number can be calculated as 
\begin{equation}
w_i=\frac{\text{Res}\mathcal{R}(z_i)}{|\text{Res}\mathcal{R}(z_i)}|=\text{Sgn}[\text{Res}\mathcal{R}(z_i)],
\end{equation}
where, $\text{Sgn}(x)$ is the sign function which
is zero for $x=0$ and for any real $x$, returns its sign. From the Cauchy-Goursat theorem, it is possible to calculate the integral of $\mathcal{R}(z)$ along an exterior contour which is over the interior contours that enclose the singular points. Mathematically,
\begin{equation}
\oint_C\mathcal{R}(z)dz=\sum_i\oint_{C_i}\mathcal{R}(z)dz.
\end{equation}
This permits the calculation of the topological number as 
\begin{equation}
W=\sum_i w_i.
\end{equation}
Various black hole systems have been studied using this approach \cite{TranNHung:2023nmb,Li:2023ppc}. The results are found to be consistent with the results obtained by the topological current method \cite{Fang:2022rsb}.

In this work, we study the thermodynamic topology of 4D Euler-Heisenberg-AdS black hole using the generalized off-shell free energy. The winding number as well as the topological numbers are calculated to analyse the local and global topological properties. We mainly focus on how different parameters of aforementioned black hole system affects the thermodynamic topology. Then, we extend our work to the higher-order QED corrected Euler-Heisenberg-AdS black hole and study the impact on the higher order correction in the thermodynamic topology. We perform our analysis in two ensembles: fixed charged (canonical) and fixed potential (grand canonical) ensemble. 

It is to be noted that the thermodynamic topology of 4D Euler-Heisenberg AdS black hole in fixed charged ensemble has been previously studied in  \cite{Alipour:2023uzo} using Duan's topological $\phi$-mapping theory. In this work, we have used a different method namely the generalized off-shell free energy method. Moreover, we conduct our study in not only the fixed charged ensemble but also the fixed potential ensemble. By doing this, we want to see how a change of ensemble affects the thermodynamic topology of this black hole. In addition to this, we also consider higher order QED corrected version of 4D EHAdS black hole. Our aim is to understand whether the higher order QED corrections changes the thermodynamic topology of the black hole or not.

This paper is organized as follows. We begin with the study of 4D Euler-Heisenberg-AdS black hole in canonical ensemble in \autoref{The Euler-Heisenberg-AdS black hole in canonical ensemble}. This is followed by, \autoref{The higher order Euler-Heisenberg-AdS black hole in canonical ensemble} in which we study the higher-order QED corrected Euler-Heisenberg-AdS black hole in canonical ensemble. Next, in \autoref{The Euler-Heisenberg-AdS black hole in grand canonical ensemble} and \autoref{The higher order Euler-Heisenberg-AdS black hole in grand canonical ensemble}, we study the Euler-Heisenberg-AdS black hole and the higher-order QED corrected Euler-Heisenberg-AdS black hole in the grand canonical ensemble respectively. Finally, we summarized our results in the \autoref{Conclusion}.
\section{4D Euler-Heisenberg-AdS black hole in canonical ensemble}
\label{The Euler-Heisenberg-AdS black hole in canonical ensemble}
The 4D Euler-Heisenberg-AdS (EHAdS) black hole is a solution of the Einstein-Maxwell equations which has a negative cosmological constant and a nonlinear electromagnetic field \cite{Salazar:1987ap,Magos:2020ykt} The nonlinear electrodynamic field arises from quantum electrodynamics corrections to the Maxwell theory and it is described by the Euler-Heisenberg Lagrangian. The metric of a static and spherically symmetric 4D EHAdS black hole is given by

\begin{equation}
ds^2 = -f(r) dt^2 + \frac{dr^2}{f(r)} + r^2 d\Omega^2.
\end{equation}
Here,
\begin{equation}
\label{Eq:EHBHFr}
f(r)= 1-\frac{2M}{r}+\frac{Q^2}{r^2}-\frac{\Lambda r^2}{3}-\frac{aQ^4}{20r^6},
\end{equation}
where $M$ is the mass, $Q$ is the electric charge, $\Lambda$ is the cosmological constant, and $a$ is the Euler-Heisenberg (EH) parameter. The parameter $a$ is proportional to the square of the fine structure constant $\alpha$ and inversely proportional to the fourth power of the electron mass $m$:
\begin{equation}
\label{Eq:EH_Parameter}
a=\frac{8\alpha^2}{45m^4},
\end{equation}
where we choose $c=\hbar=1$. The EH parameter measures the strength of the quantum electrodynamics (QED) correction to the Maxwell field.

The thermodynamics and phase transitions of this black hole have already been studied in a number of remarkable works \cite{Magos:2020ykt,Dai:2022mko}. It has been found that the EH parameter $a$ affects the thermodynamics and phase transition of the Euler-Heisenberg-AdS black hole. Depending on the value of a, the black hole can exhibit interesting phase transition behaviors. Such as, for $a<0$, the black hole can undergo a first-order phase transition between small and large black holes, similar to the van der Waals system. For $0 \leq a \leq \frac{32}{7}Q^2$, there is reentrant phase transition. For $a>\frac{32}{7}Q^2$, there is no phase transition \cite{Ye:2022uuj}. 

The cosmological constant can be represented in terms of pressure as $\Lambda=-8\pi P$. With this definition of pressure the relevant thermodynamic parameters such as entropy, energy and temperature of the black hole are respectively given as
\begin{equation}
S=\pi r_+^2,
\end{equation}
\begin{equation}
\label{Eq:EHBHMass}
E=M=\frac{r_+}{2}+\frac{4}{3} \pi  P r_+^3+\frac{Q^2}{2 r_+}-\frac{a Q^4}{40 r_+^5},
\end{equation}
and
\begin{equation}
\label{Eq:EHBHTemp}
T=\frac{a Q^4+32 \pi  P r_+^8-4 Q^2 r_+^4+4 r_+^6}{16 \pi  r_+^7}.
\end{equation}
The generalized free energy is given as 
\begin{equation}
\begin{aligned}
\mathcal{F}&=E-\frac{S}{\tau} \\
&=\frac{4}{3} \pi  P r_+^3+\frac{Q^2}{2 r_+}-\frac{a Q^4}{40 r_+^5}-\frac{\pi  r_+^2}{\tau }+\frac{r_+}{2},
\end{aligned}
\end{equation}
where, $\tau$ is the inverse of temperature. For the Duan's $\phi$ mapping theory, the vector field is defind as 
\begin{equation}
\label{Eq:Vector_Field}
\phi=(\partial_{r_+}\mathcal{F},-\cot\Theta\csc\Theta),
\end{equation}
where, $\Theta$ is added here for convenience and $0\le\Theta\le\pi$. Using $\partial_{r_+}\mathcal{F}=0$, we calculate the curve of zero points of $\phi$:
\begin{equation}
\label{Eq:General_Tau}
\tau=\frac{16 \pi  r_+^7}{a Q^4+32 \pi  P r_+^8-4 Q^2 r_+^4+4 r_+^6}.
\end{equation}
Now, from \eqref{Eq:GeneralComplexFunction}, we define the complex function as 
\begin{equation}
\label{Eq:REHBH}
\mathcal{R}_{EH}(z)=\frac{a Q^4+32 \pi  P z^8-4 Q^2 z^4+4 z^6}{a Q^4 \tau +32 \pi  P \tau  z^8-4 Q^2 \tau  z^4+4 \tau  z^6-16 \pi  z^7}.
\end{equation}
We take the denominator of the equation and define a polynomial function $\mathcal{A}(z)$ as
\begin{equation}
\label{Eq:General_Polynimial_Function}
\mathcal{A}(z)=a Q^4 \tau +32 \pi  P \tau  z^8-4 Q^2 \tau  z^4+4 \tau  z^6-16 \pi  z^7.
\end{equation}
Using $\mathcal{A}(z)$, we can calculate the poles of $\mathcal{R}(z)_{EH}$.
\subsection{For negative $a$}
\label{EHBH negative $a$}
\begin{figure}[ht]
	\centering
		\begin{subfigure}{0.5\textwidth}
			\centering
			\includegraphics[width=0.9\linewidth]{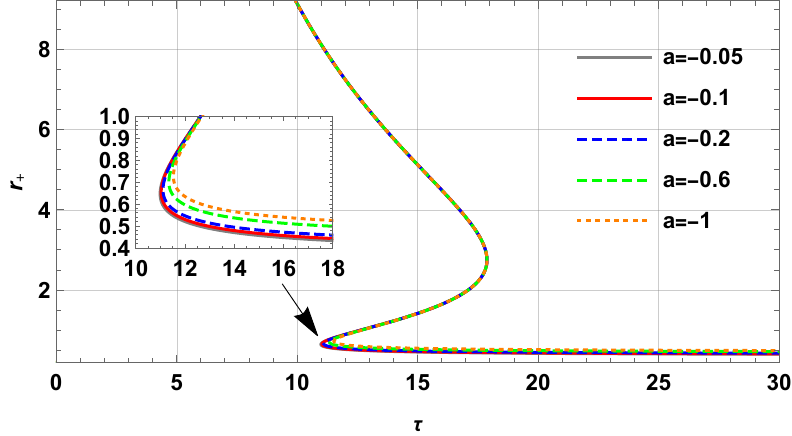}
			\caption{Zero points of $\phi$ for different negative values of $a$. Here $P=0.005$ and $Q=0.36$.}
			\label{Fig:EHBH_Negative_a_(a_Variation)}
		\end{subfigure}%
		\begin{subfigure}{0.5\textwidth}
			\centering
			\includegraphics[width=0.9\linewidth]{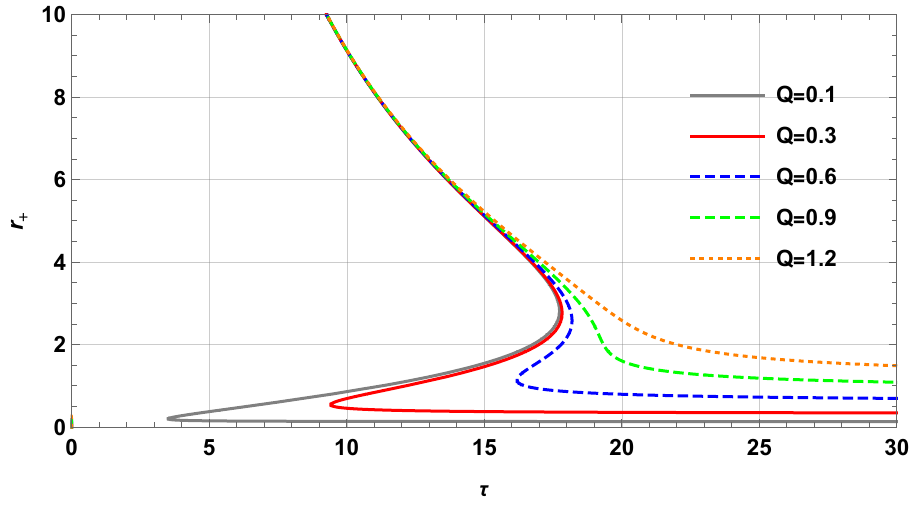}
			\caption{Zero points of $\phi$ for different values of $Q$. Here $a=-0.1$ and $P=0.005$}
			\label{Fig:EHBH_Negative_a_(Q_Variation)}
		\end{subfigure} \\
		\begin{subfigure}{0.5\textwidth}
			\centering
			\includegraphics[width=0.9\linewidth]{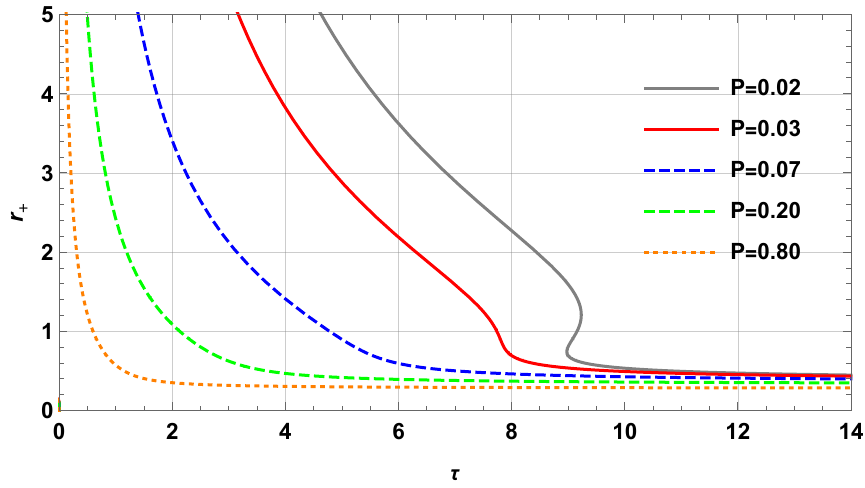}
			\caption{Zero points of $\phi$ for different values of $P$. Here $a=-0.1$ and $Q=0.36$}
			\label{Fig:EHBH_Negative_a_(P_Variation)}
		\end{subfigure}%
	\caption{Zero points of $\phi$ in $r_+$ vs $\tau$ plane with different parameter variations for EHAdS black hole in canonical ensemble (for negative $a$).}
	\label{Fig:EHBH_Tau_Plot_All_Negative}
	\end{figure}

We first consider the case of the negative EH parameter $(a<0)$. For such a case, variations of $\tau(r_+)$ for different parameters are shown in \autoref{Fig:EHBH_Tau_Plot_All_Negative}.
For a set of parameters $(a,P,Q)=(-0.1,0.005,0.36)$, we proceed to calculate the poles of \eqref{Eq:REHBH} by substituting these values in the polynomial function \eqref{Eq:General_Polynimial_Function}. This provides
\begin{equation}
\mathcal{A}(z)=-0.00167962 \tau +0.502655 \tau  z^8+4 \tau  z^6-0.5184 \tau  z^4-16 \pi  z^7.
\end{equation}
A graphical representation in \autoref{Fig:Polynomial_Plot_a_Negative_0dot5_P_0dot005,Q_0dot36} shows the roots of the polynomial function $\mathcal{A}(z)$ for different values of $\tau$. For example, with $\tau=15$, we have three positive real roots of $\mathcal{A}(z)$ which are $z_1=0.466388, ~ z_2=1.41858, ~ z_3=5.12015$.
\begin{figure}[h!]
	\centerline{
	\includegraphics[scale=0.65]{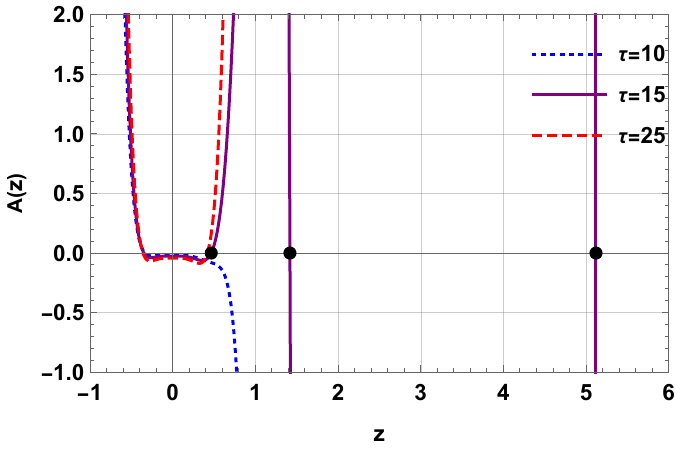}}
	\caption	{Plot of the polynomial function $\mathcal{A}(z)$ for $\tau=10,~15,\text{and}~25$. The black dots represents the roots for $\tau=15$.}	\label{Fig:Polynomial_Plot_a_Negative_0dot5_P_0dot005,Q_0dot36}
	\end{figure}
Using residue theorem we can calculate the winding numbers. For $z_1,~z_2$ and $z_3$ the winding numbers are
\begin{equation}
w_1=+1, \quad w_2=-1, \quad w_3=+1.
\end{equation}
The topological number is hence $W=w_1+w_2+w_3=+1$. In the same parameter configuration, for $\tau=10$ and $\tau=25$ we find $w=+1$ for each value of $\tau$. This imply that the topological number, $W$, is also $+1$.
We also calculate the winding number as well as the topological number for different variations of parameters and the results are shown in \autoref{EHBH_Negative_a}. The results shows that the topological number $W=+1$ is independent of the parameter variation.
\begin{table}[h!]
\centering
\begin{tabular}{|cccccc|}
\hline
\multicolumn{6}{|c|}{EHAdS black hole}                                                                                                                                                                                                                                                                                       \\ \hline
\multicolumn{3}{|c|}{Parameters}                                                                                                       & \multicolumn{1}{c|}{\multirow{2}{*}{$\tau$}} & \multicolumn{1}{c|}{\multirow{2}{*}{Winding number}}                                        & \multirow{2}{*}{Topological number} \\ \cline{1-3}
\multicolumn{1}{|c|}{P}                      & \multicolumn{1}{c|}{Q}                     & \multicolumn{1}{c|}{a}                     & \multicolumn{1}{c|}{}                        & \multicolumn{1}{c|}{}                                                                       &                                     \\ \hline
\multicolumn{1}{|c|}{\multirow{3}{*}{0.005}} & \multicolumn{1}{c|}{\multirow{3}{*}{0.36}} & \multicolumn{1}{c|}{\multirow{3}{*}{-0.1}} & \multicolumn{1}{c|}{10}                      & \multicolumn{1}{c|}{$w_1=+1$}                                                               & $W=+1$                              \\ \cline{4-6} 
\multicolumn{1}{|c|}{}                       & \multicolumn{1}{c|}{}                      & \multicolumn{1}{c|}{}                      & \multicolumn{1}{c|}{15}                      & \multicolumn{1}{c|}{\begin{tabular}[c]{@{}c@{}}$w_1=+1$\\ $w_2=-1$\\ $w_3=+1$\end{tabular}} & $W=+1$                              \\ \cline{4-6} 
\multicolumn{1}{|c|}{}                       & \multicolumn{1}{c|}{}                      & \multicolumn{1}{c|}{}                      & \multicolumn{1}{c|}{25}                      & \multicolumn{1}{c|}{$w_1=+1$}                                                               & $W=+1$                              \\ \hline
\multicolumn{1}{|c|}{\multirow{3}{*}{0.005}} & \multicolumn{1}{c|}{\multirow{3}{*}{0.36}} & \multicolumn{1}{c|}{\multirow{3}{*}{-0.6}} & \multicolumn{1}{c|}{10}                      & \multicolumn{1}{c|}{$w_1=+1$}                                                               & $W=+1$                              \\ \cline{4-6} 
\multicolumn{1}{|c|}{}                       & \multicolumn{1}{c|}{}                      & \multicolumn{1}{c|}{}                      & \multicolumn{1}{c|}{15}                      & \multicolumn{1}{c|}{\begin{tabular}[c]{@{}c@{}}$w_1=+1$\\ $w_2=-1$\\ $w_3=+1$\end{tabular}} & $W=+1$                              \\ \cline{4-6} 
\multicolumn{1}{|c|}{}                       & \multicolumn{1}{c|}{}                      & \multicolumn{1}{c|}{}                      & \multicolumn{1}{c|}{25}                      & \multicolumn{1}{c|}{$w_1=+1$}                                                               & $W=+1$                              \\ \hline
\multicolumn{1}{|c|}{\multirow{3}{*}{0.005}} & \multicolumn{1}{c|}{\multirow{3}{*}{0.1}}  & \multicolumn{1}{c|}{\multirow{3}{*}{-0.1}} & \multicolumn{1}{c|}{5}                       & \multicolumn{1}{c|}{\begin{tabular}[c]{@{}c@{}}$w_1=+1$\\ $w_2=-1$\\ $w_3=+1$\end{tabular}} & $W=+1$                              \\ \cline{4-6} 
\multicolumn{1}{|c|}{}                       & \multicolumn{1}{c|}{}                      & \multicolumn{1}{c|}{}                      & \multicolumn{1}{c|}{10}                      & \multicolumn{1}{c|}{\begin{tabular}[c]{@{}c@{}}$w_1=+1$\\ $w_2=-1$\\ $w_3=+1$\end{tabular}} & $W=+1$                              \\ \cline{4-6} 
\multicolumn{1}{|c|}{}                       & \multicolumn{1}{c|}{}                      & \multicolumn{1}{c|}{}                      & \multicolumn{1}{c|}{15}                      & \multicolumn{1}{c|}{\begin{tabular}[c]{@{}c@{}}$w_1=+1$\\ $w_2=-1$\\ $w_3=+1$\end{tabular}} & $W=+1$                              \\ \hline
\multicolumn{1}{|c|}{\multirow{3}{*}{0.005}} & \multicolumn{1}{c|}{\multirow{3}{*}{0.9}}  & \multicolumn{1}{c|}{\multirow{3}{*}{-0.1}} & \multicolumn{1}{c|}{10}                      & \multicolumn{1}{c|}{$w_1=+1$}                                                               & $W=+1$                              \\ \cline{4-6} 
\multicolumn{1}{|c|}{}                       & \multicolumn{1}{c|}{}                      & \multicolumn{1}{c|}{}                      & \multicolumn{1}{c|}{15}                      & \multicolumn{1}{c|}{$w_1=+1$}                                                               & $W=+1$                              \\ \cline{4-6} 
\multicolumn{1}{|c|}{}                       & \multicolumn{1}{c|}{}                      & \multicolumn{1}{c|}{}                      & \multicolumn{1}{c|}{20}                      & \multicolumn{1}{c|}{$w_1=+1$}                                                               & $W=+1$                              \\ \hline
\multicolumn{1}{|c|}{\multirow{3}{*}{0.2}}   & \multicolumn{1}{c|}{\multirow{3}{*}{0.36}} & \multicolumn{1}{c|}{\multirow{3}{*}{-0.1}} & \multicolumn{1}{c|}{4}                       & \multicolumn{1}{c|}{$w_1=+1$}                                                               & $W=+1$                              \\ \cline{4-6} 
\multicolumn{1}{|c|}{}                       & \multicolumn{1}{c|}{}                      & \multicolumn{1}{c|}{}                      & \multicolumn{1}{c|}{8}                       & \multicolumn{1}{c|}{$w_1=+1$}                                                               & $W=+1$                              \\ \cline{4-6} 
\multicolumn{1}{|c|}{}                       & \multicolumn{1}{c|}{}                      & \multicolumn{1}{c|}{}                      & \multicolumn{1}{c|}{12}                      & \multicolumn{1}{c|}{$w_1=+1$}                                                               & $W=+1$                              \\ \hline
\end{tabular}
\caption{Winding numbers and Topological numbers for EHAdS black hole in canonical ensemble for negative $a$}
\label{EHBH_Negative_a}
\end{table}

\subsection{For positive $a$}
\begin{figure}[ht]
	\centering
		\begin{subfigure}{0.5\textwidth}
			\centering
			\includegraphics[width=0.9\linewidth]{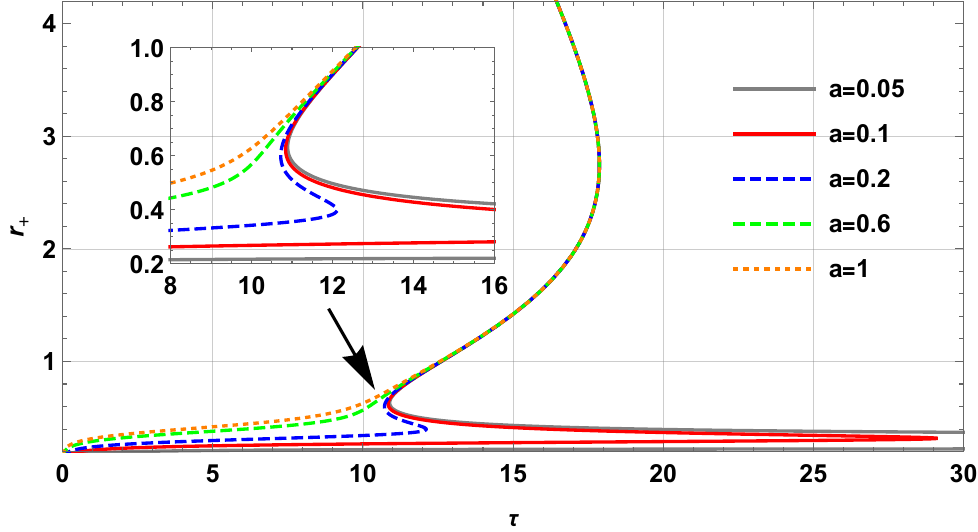}
			\caption{Zero points of $\phi$ for different positive values of $a$. Here $P=0.005$ and $Q=0.36$.}
			\label{Fig:EHBH_Positive_a_(a_Variation)}
		\end{subfigure}%
		\begin{subfigure}{0.5\textwidth}
			\centering
			\includegraphics[width=0.9\linewidth]{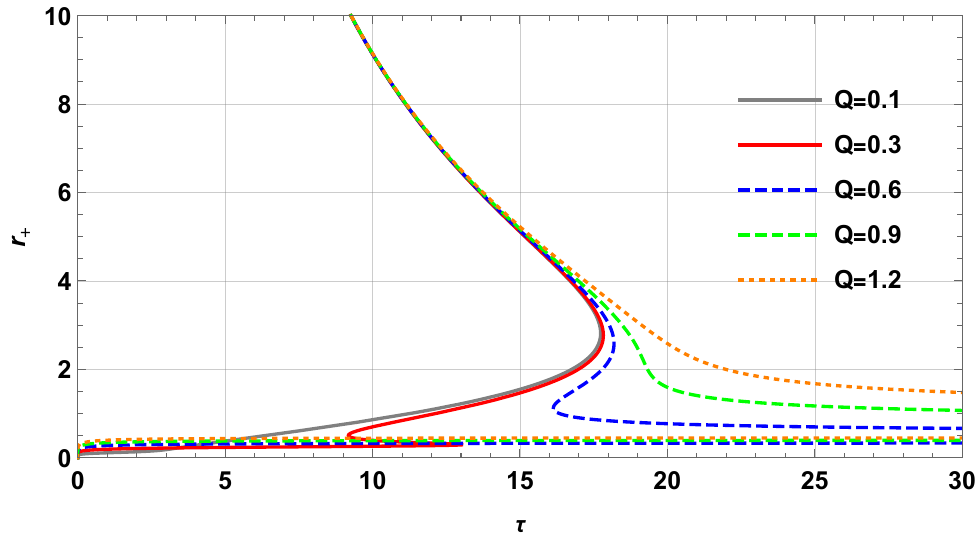}
			\caption{Zero points of $\phi$ for different values of $Q$. Here $a=0.1$ and $P=0.005$}
			\label{Fig:EHBH_Positive_a_(Q_Variation)}
		\end{subfigure} \\
		\begin{subfigure}{0.5\textwidth}
			\centering
			\includegraphics[width=0.9\linewidth]{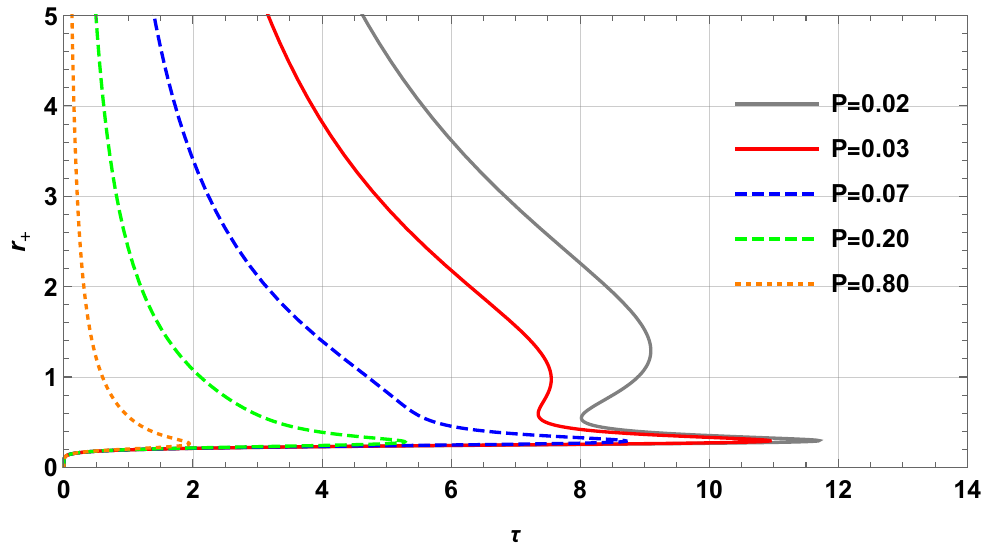}
			\caption{Zero points of $\phi$ for different values of $P$. Here $a=0.1$ and $Q=0.36$}
			\label{Fig:EHBH_Positive_a_(P_Variation)}
		\end{subfigure}%
	\caption{Zero points of $\phi$ in $r_+$ vs $\tau$ plane with different parameter variations for EHAdS black hole in canonical ensemble (positive $a$).}
	\label{Fig:EHBH_Tau_Plot_All_Positive}
	\end{figure}

\begin{figure}[ht]
	\centering
		\begin{subfigure}{0.5\textwidth}
			\centering
			\includegraphics[width=0.9\linewidth]{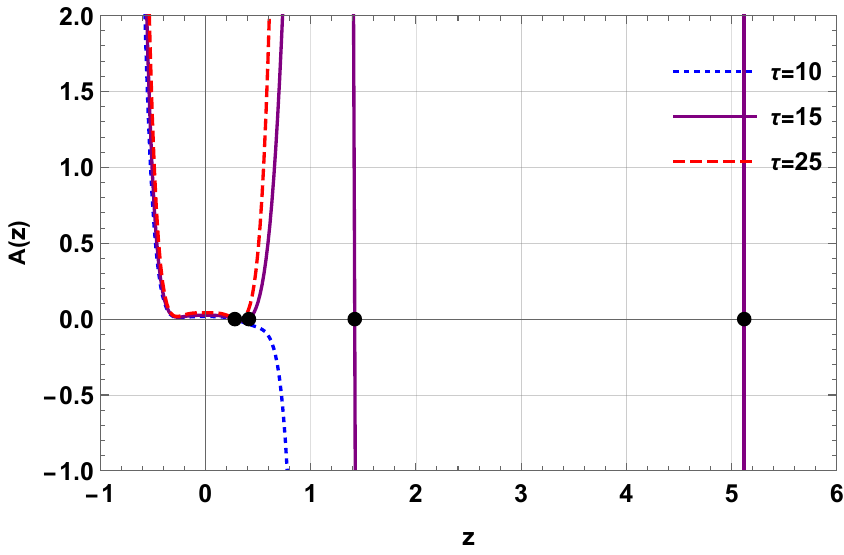}
			\caption{All four roots of $\mathcal{A}(z)$ for $\tau=15$}
			\label{Fig:Polynomial_Plot_a_Positive_0dot1_P_0dot005,Q_0dot36}
		\end{subfigure}%
		\begin{subfigure}{0.5\textwidth}
			\centering
			\includegraphics[width=0.9\linewidth]{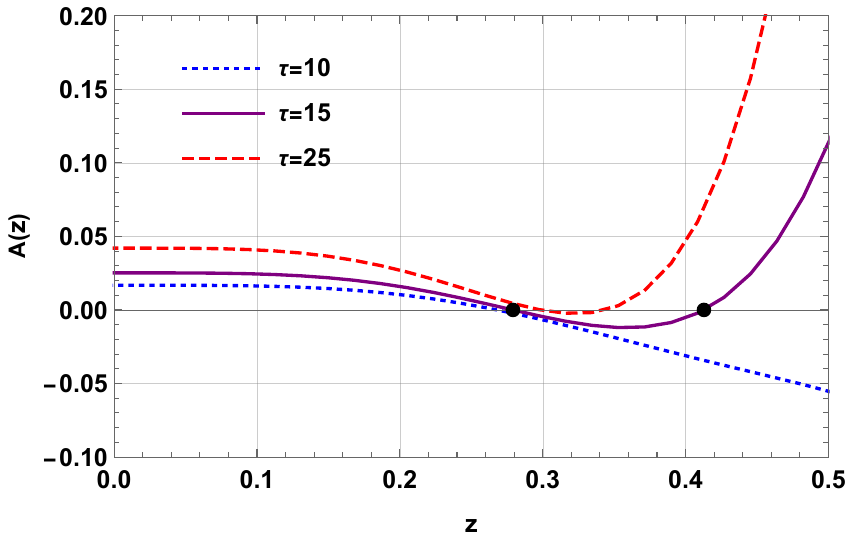}
			\caption{Enlarged view of the two smaller positive roots of $\mathcal{A}(z)$ for $\tau=15$}
			\label{Fig:Polynomial_Plot_a_Positive_0dot1_P_0dot005,Q_0dot36_Small_Region}
		\end{subfigure} \\
	\caption{Plot of the polynomial function $\mathcal{A}(z)$ for $\tau=10,~15,\text{and}~25$. The black dots represents the roots for $\tau=15$.}
	\label{Fig:EHBH_Poly_Plot_All_Positive}
	\end{figure}

Here, we consider the EH parameter $a$ to be positive, i.e., $a>0$. For this case, the zero points of $\phi$ are shown in \autoref{Fig:EHBH_Tau_Plot_All_Positive} for a range different constant parameters. The figures are plotted using \eqref{Eq:General_Tau}. From the figure we observe either $2$ or $4$ different branches of the $\tau(r_+)$ curve.
Now, we choose a parameter combination $(a,P,Q)=(0.1,0.005,0.36)$ such that
\begin{equation}
\label{Eq:Positive_Polynomial}
\mathcal{A}(z)=0.00167962 \tau +0.502655 \tau  z^8+4 \tau  z^6-0.5184 \tau  z^4-16 \pi  z^7.
\end{equation} 
The plot of $\mathcal{A}(z)$ is shown in \autoref{Fig:EHBH_Poly_Plot_All_Positive}. 
Using \eqref{Eq:Positive_Polynomial} we can calculate the roots for fixed $\tau$. For $\tau=15$ the roots are shown in \autoref{Fig:EHBH_Poly_Plot_All_Positive} as black dots which are are $z_1=0.279175,~z_2=0.412857,~z_3=1.41885,~\text{and}~z_4=5.12015$. Now using \eqref{Eq:REHBH} and following the Residue theorem, for each positive real roots, we calculate the winding numbers
\begin{equation}
w_1=-1,~w_2=+1,~w_3=-1,~ \text{and} ~w_4=+1.
\end{equation}
Clearly, here the topological number is $W=w_1+w_2+w_3+w_4=-1+1-1+1=0$. To understand the impact of different parameters in the winding numbers and the topological number, we perform our study for a variety of parameter values. The results are summarized in the \autoref{EHBH_Positive_a}.
\begin{table}[ht]
\centering
\begin{tabular}{|cccccc|}
\hline
\multicolumn{6}{|c|}{EHAdS black hole}                                                                                                                                                                                                                                                                                                 \\ \hline
\multicolumn{3}{|c|}{Parameters}                                                                                                      & \multicolumn{1}{c|}{\multirow{2}{*}{$\tau$}} & \multicolumn{1}{c|}{\multirow{2}{*}{Winding number}}                                                   & \multirow{2}{*}{Topological number} \\ \cline{1-3}
\multicolumn{1}{|c|}{P}                      & \multicolumn{1}{c|}{Q}                     & \multicolumn{1}{c|}{a}                    & \multicolumn{1}{c|}{}                        & \multicolumn{1}{c|}{}                                                                                  &                                     \\ \hline
\multicolumn{1}{|c|}{\multirow{3}{*}{0.005}} & \multicolumn{1}{c|}{\multirow{3}{*}{0.36}} & \multicolumn{1}{c|}{\multirow{3}{*}{0.1}} & \multicolumn{1}{c|}{10}                      & \multicolumn{1}{c|}{\begin{tabular}[c]{@{}c@{}}$w_1=-1$\\ $w_2=+1$\end{tabular}}                       & $W=0$                               \\ \cline{4-6} 
\multicolumn{1}{|c|}{}                       & \multicolumn{1}{c|}{}                      & \multicolumn{1}{c|}{}                     & \multicolumn{1}{c|}{15}                      & \multicolumn{1}{c|}{\begin{tabular}[c]{@{}c@{}}$w_1=-1$\\ $w_2=+1$\\ $w_3=-1$\\ $w_4=+1$\end{tabular}} & $W=0$                               \\ \cline{4-6} 
\multicolumn{1}{|c|}{}                       & \multicolumn{1}{c|}{}                      & \multicolumn{1}{c|}{}                     & \multicolumn{1}{c|}{25}                      & \multicolumn{1}{c|}{\begin{tabular}[c]{@{}c@{}}$w_1=-1$\\ $w_2=+1$\end{tabular}}                       & $W=0$                               \\ \hline
\multicolumn{1}{|c|}{\multirow{3}{*}{0.005}} & \multicolumn{1}{c|}{\multirow{3}{*}{0.36}} & \multicolumn{1}{c|}{\multirow{3}{*}{0.6}} & \multicolumn{1}{c|}{10}                      & \multicolumn{1}{c|}{\begin{tabular}[c]{@{}c@{}}$w_1=-1$\\ $w_2=+1$\end{tabular}}                       & $W=0$                               \\ \cline{4-6} 
\multicolumn{1}{|c|}{}                       & \multicolumn{1}{c|}{}                      & \multicolumn{1}{c|}{}                     & \multicolumn{1}{c|}{\multirow{2}{*}{15}}     & \multicolumn{1}{c|}{\multirow{2}{*}{\begin{tabular}[c]{@{}c@{}}$w_1=-1$\\ $w_2=+1$\end{tabular}}}      & \multirow{2}{*}{$W=0$}              \\
\multicolumn{1}{|c|}{}                       & \multicolumn{1}{c|}{}                      & \multicolumn{1}{c|}{}                     & \multicolumn{1}{c|}{}                        & \multicolumn{1}{c|}{}                                                                                  &                                     \\ \hline
\multicolumn{1}{|c|}{\multirow{3}{*}{0.005}} & \multicolumn{1}{c|}{\multirow{3}{*}{0.1}}  & \multicolumn{1}{c|}{\multirow{3}{*}{0.1}} & \multicolumn{1}{c|}{5}                       & \multicolumn{1}{c|}{\begin{tabular}[c]{@{}c@{}}$w_1=-1$\\ $w_2=+1$\end{tabular}}                       & $W=0$                               \\ \cline{4-6} 
\multicolumn{1}{|c|}{}                       & \multicolumn{1}{c|}{}                      & \multicolumn{1}{c|}{}                     & \multicolumn{1}{c|}{10}                      & \multicolumn{1}{c|}{\begin{tabular}[c]{@{}c@{}}$w_1=-1$\\ $w_2=+1$\end{tabular}}                       & $W=0$                               \\ \cline{4-6} 
\multicolumn{1}{|c|}{}                       & \multicolumn{1}{c|}{}                      & \multicolumn{1}{c|}{}                     & \multicolumn{1}{c|}{15}                      & \multicolumn{1}{c|}{\begin{tabular}[c]{@{}c@{}}$w_1=-1$\\ $w_2=+1$\end{tabular}}                       & $W=0$                               \\ \hline
\multicolumn{1}{|c|}{\multirow{4}{*}{0.005}} & \multicolumn{1}{c|}{\multirow{4}{*}{0.9}}  & \multicolumn{1}{c|}{\multirow{4}{*}{0.1}} & \multicolumn{1}{c|}{5}                       & \multicolumn{1}{c|}{\begin{tabular}[c]{@{}c@{}}$w_1=-1$\\ $w_2=+1$\end{tabular}}                       & $W=0$                               \\ \cline{4-6} 
\multicolumn{1}{|c|}{}                       & \multicolumn{1}{c|}{}                      & \multicolumn{1}{c|}{}                     & \multicolumn{1}{c|}{10}                      & \multicolumn{1}{c|}{\begin{tabular}[c]{@{}c@{}}$w_1=-1$\\ $w_2=+1$\end{tabular}}                       & $W=0$                               \\ \cline{4-6} 
\multicolumn{1}{|c|}{}                       & \multicolumn{1}{c|}{}                      & \multicolumn{1}{c|}{}                     & \multicolumn{1}{c|}{15}                      & \multicolumn{1}{c|}{\begin{tabular}[c]{@{}c@{}}$w_1=-1$\\ $w_2=+1$\end{tabular}}                       & $W=0$                               \\ \cline{4-6} 
\multicolumn{1}{|c|}{}                       & \multicolumn{1}{c|}{}                      & \multicolumn{1}{c|}{}                     & \multicolumn{1}{c|}{20}                      & \multicolumn{1}{c|}{\begin{tabular}[c]{@{}c@{}}$w_1=-1$\\ $w_2=+1$\end{tabular}}                       & $W=0$                               \\ \hline
\multicolumn{1}{|c|}{\multirow{3}{*}{0.2}}   & \multicolumn{1}{c|}{\multirow{3}{*}{0.36}} & \multicolumn{1}{c|}{\multirow{3}{*}{0.1}} & \multicolumn{1}{c|}{1}                       & \multicolumn{1}{c|}{\begin{tabular}[c]{@{}c@{}}$w_1=-1$\\ $w_2=+1$\end{tabular}}                       & $W=0$                               \\ \cline{4-6} 
\multicolumn{1}{|c|}{}                       & \multicolumn{1}{c|}{}                      & \multicolumn{1}{c|}{}                     & \multicolumn{1}{c|}{3}                       & \multicolumn{1}{c|}{\begin{tabular}[c]{@{}c@{}}$w_1=-1$\\ $w_2=+1$\end{tabular}}                       & $W=0$                               \\ \cline{4-6} 
\multicolumn{1}{|c|}{}                       & \multicolumn{1}{c|}{}                      & \multicolumn{1}{c|}{}                     & \multicolumn{1}{c|}{5}                       & \multicolumn{1}{c|}{\begin{tabular}[c]{@{}c@{}}$w_1=-1$\\ $w_2=+1$\end{tabular}}                       & $W=0$                              \\ \hline
\end{tabular}
\caption{Winding numbers and Topological numbers for EHAdS black hole in canonical ensemble for positive $a$}
\label{EHBH_Positive_a}
\end{table}

\section{Higher-order QED corrected Euler-Heisenberg-AdS black hole in canonical ensemble}
\label{The higher order Euler-Heisenberg-AdS black hole in canonical ensemble}
In this section we extend our work to the higher-order QED corrected Euler-Heisenberg-AdS black hole \cite{Ye:2022uuj,Li:2021ygi}. We study the system by treating it as thermodynamic topological defects and study its topological properties. The higher-order Euler-Heisenberg-AdS black hole is obtained by the higher order QED correction of the EHAdS black hole. In this higher order correction, the metric potential, mass, temperature are modified as 
\begin{equation}
\label{Eq:HOEHBH_Fr}
f^\prime(r)=f(r)+\frac{\beta a^2 Q^2}{36 r^{10}},
\end{equation}
\begin{equation}
\label{Eq:HOEHBH_Mass}
M^\prime=M+\frac{\beta a^2 Q^6}{72 r_+^9},
\end{equation}
\begin{equation}
\label{Eq:HOEHBH_Temperature}
T^\prime=T-\frac{\beta a^2 Q^6}{16 \pi r_+^{11}},
\end{equation}
where, $f(r)$, $M$ and $T$ are given in \eqref{Eq:EHBHFr}, \eqref{Eq:EHBHMass} and \eqref{Eq:EHBHTemp}. Here, we have an additional parameter $\beta$ which appears as a consequence of the higher order QED correction. The equations \eqref{Eq:HOEHBH_Fr}, \eqref{Eq:HOEHBH_Mass}, and \eqref{Eq:HOEHBH_Temperature} reduces to its corresponding EHAdS black hole thermodynamic parameters when $\beta \rightarrow 0$.
Now, the generalized free energy is given as
\begin{equation}
\mathcal{F}=\frac{a^2 \beta  Q^6}{72 r_+^9}+\frac{-3 a Q^4+160 \pi  P r_+^8+60 Q^2 r_+^4+60 r_+^6}{120 r_+^5}-\frac{\pi  r_+^2}{\tau }
\end{equation}
The vector field $\phi$ is defined by \eqref{Eq:Vector_Field} and the zero points $\phi$ are given by
\begin{equation}
\label{Eq:HOEH_Tau}
\tau=\frac{16 \pi  r_+^{11}}{-a^2 \beta  Q^6+a Q^4 r_+^4+32 \pi  P r_+^{12}-4 Q^2 r_+^8+4 r_+^{10}}.
\end{equation}
Using \eqref{Eq:GeneralComplexFunction} we define the complex function
\begin{equation}
\label{Eq:HOEH_Complex_Function}
\mathcal{R}_{HOEH}(z)=\frac{-a^2 \beta  Q^6+a Q^4 z^4+32 \pi  P z^{12}-4 Q^2 z^8+4 z^{10}}{-a^2 \beta  Q^6 \tau +a Q^4 \tau  z^4+32 \pi  P \tau  z^{12}-4 Q^2 \tau  z^8+4 \tau  z^{10}-16 \pi  z^{11}}.
\end{equation}
Denominator of \eqref{Eq:HOEH_Complex_Function} is considered as a polynomial function
\begin{equation}
\label{Eq:HOEH_Polynomial}
\mathcal{A}(z)=-a^2 \beta  Q^6 \tau +a Q^4 \tau  z^4+32 \pi  P \tau  z^{12}-4 Q^2 \tau  z^8+4 \tau  z^{10}-16 \pi  z^{11}.
\end{equation}
The roots of $\mathcal{A}(z)$ will give the poles of $\mathcal{R}_{HOEH}(z)$.
\subsection{For negative $a$}
\begin{figure}[ht]
	\centering
		\begin{subfigure}{0.5\textwidth}
			\centering			\includegraphics[width=0.9\linewidth]{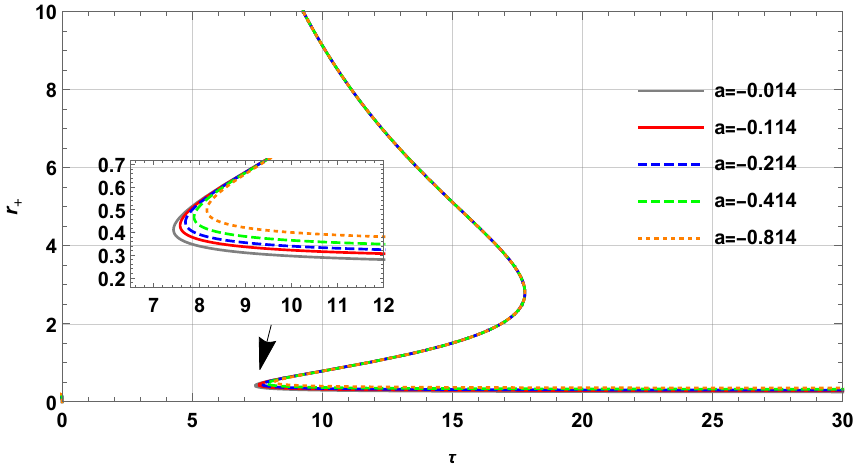}
			\caption{Zero points of $\phi$ for different negative values of $a$. Here $\beta=0.11$, $P=0.005$ and $Q=0.234$.}		\label{Fig:HOEHBH_Negative_a_(a_Variation)}
		\end{subfigure}%
		\begin{subfigure}{0.5\textwidth}
			\centering
		\includegraphics[width=0.9\linewidth]{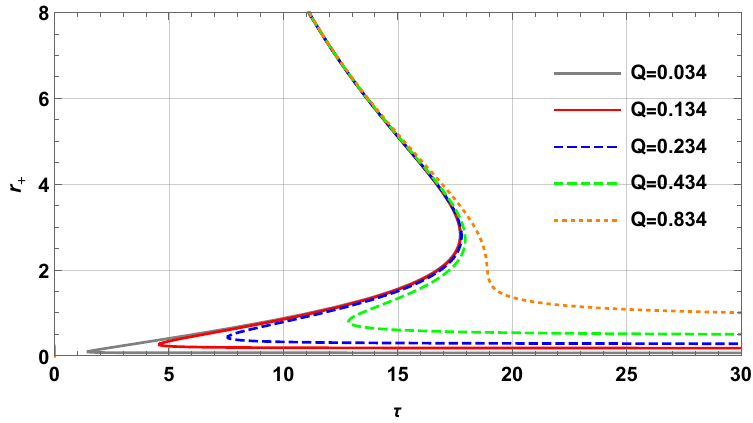}
			\caption{Zero points of $\phi$ for different values of $Q$. Here $\beta=0.11$, $a=-0.114$ and $P=0.005$}			\label{Fig:HOEHBH_Negative_a_(Q_Variation)}
		\end{subfigure} \\
		\begin{subfigure}{0.5\textwidth}
			\centering			\includegraphics[width=0.9\linewidth]{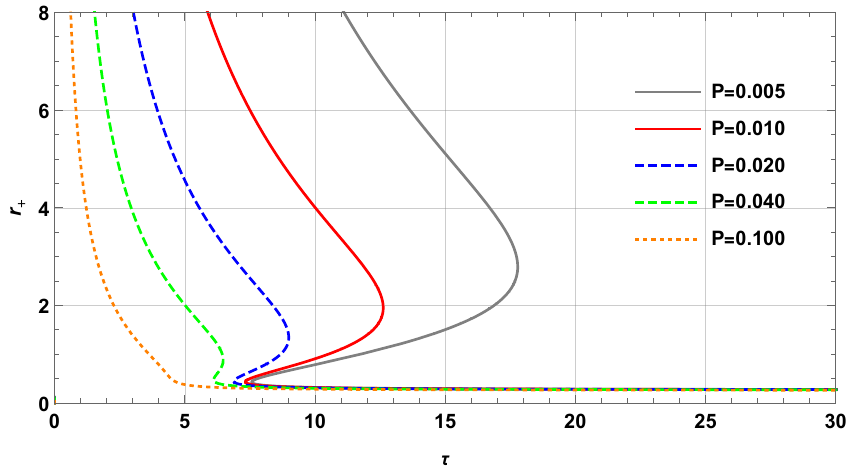}
			\caption{Zero points of $\phi$ for different values of $P$. Here $\beta=0.11$, $a=-0.114$ and $Q=0.36$}			\label{Fig:HOEHBH_Negative_a_(P_Variation)}
		\end{subfigure}%
	\caption{$\tau$ vs $r_+$ plot with different parameter variations for higher-order QED corrected EHAdS black hole in canonical ensemble (negative $a$).}
	\label{Fig:HOEHBH_Tau_Plot_All_Negative}
	\end{figure}

Here, we study the system with $\beta=0.11$ and for $a<0$.
The variation of $\tau(r_+)$ is shown in \autoref{Fig:HOEHBH_Tau_Plot_All_Negative}. From these figure we see that $\tau(r_+)$ can have either $1$ or $3$ different branches depending on the parameters $P$, $Q$, $a$ and $\beta$.
The polynomial function is calculated from \eqref{Eq:HOEH_Polynomial} and for $(P,Q,a,\beta)=(0.005,0.234,-0.114,0.11)$ it is given as
\begin{equation}
\mathcal{A}(z)=-2.346915927219771 \times 10^ {-7}\tau +0.502655 \tau  z^{12}+4 \tau  z^{10}-0.219024 \tau  z^8-0.000341797 \tau  z^4-16 \pi  z^{11}
\end{equation}
The roots for different values of $\tau$ is shown in \autoref{Fig:Polynomial_Plot_HO_a_Negative_0dot114_P_0dot005,Q_0dot234_beta_0.11}. 
We choose $\tau=15$ and find three real positive roots as $z_1=0.296051$, $z_2=1.50412$ and $z_3=5.11379$. These roots are shown in \autoref{Fig:Polynomial_Plot_HO_a_Negative_0dot114_P_0dot005,Q_0dot234_beta_0.11} as black dots.
\begin{figure}[h!]
	\centerline{
	\includegraphics[scale=0.5]{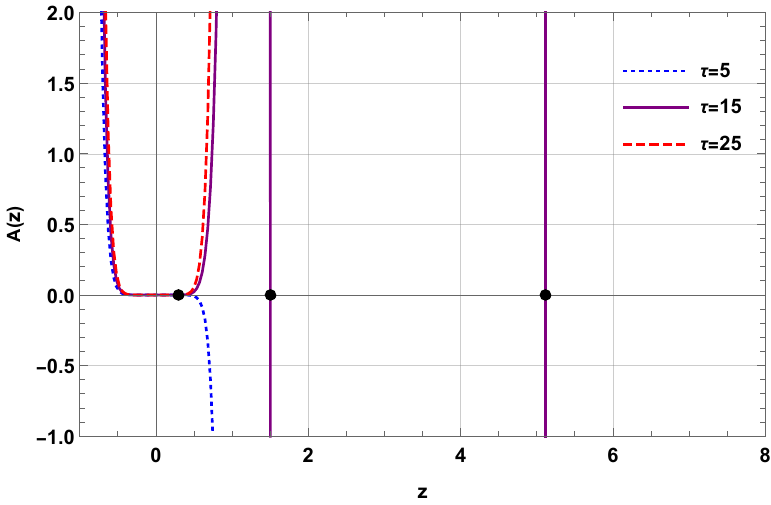}}
	\caption	{Plot of the polynomial function $\mathcal{A}(z)$ for $\tau=5,~15,\text{and}~25$. The black dots represents the roots for $\tau=15$.}	\label{Fig:Polynomial_Plot_HO_a_Negative_0dot114_P_0dot005,Q_0dot234_beta_0.11}
	\end{figure}
Now using the residue we calculate their corresponding winding numbers which are $w_1=+1$, $w_2=-1$, and $w_3=+1$. The topological number is hence $W=+1-1+1=+1$. From \autoref{Fig:HOEHBH_Negative_a_(a_Variation)} we see that the winding numbers are on the different branches of $\tau(r_+)$ curve and $w_1=w_3=+1$ corresponds to the stable black hole region whereas the $w_2=-1$ corresponds to the unstable black hole region. We have calculated the winding numbers for different values of the parameters and the results are shown in the \autoref{HOEHBH_Negative_a}.	
\begin{table}[ht]
\centering
\begin{tabular}{|cccccc|}
\hline
\multicolumn{6}{|c|}{Higher order EHAdS black hole ($\beta = 0.11$)}                                                                                                                                                                                                                                                            \\ \hline
\multicolumn{3}{|c|}{Parameters}                                                                                                          & \multicolumn{1}{c|}{\multirow{2}{*}{$\tau$}} & \multicolumn{1}{c|}{\multirow{2}{*}{Winding number}}                                        & \multirow{2}{*}{Topological number} \\ \cline{1-3}
\multicolumn{1}{|c|}{P}                      & \multicolumn{1}{c|}{Q}                      & \multicolumn{1}{c|}{a}                       & \multicolumn{1}{c|}{}                        & \multicolumn{1}{c|}{}                                                                       &                                     \\ \hline
\multicolumn{1}{|c|}{\multirow{3}{*}{0.005}} & \multicolumn{1}{c|}{\multirow{3}{*}{0.234}} & \multicolumn{1}{c|}{\multirow{3}{*}{-0.114}} & \multicolumn{1}{c|}{5}                       & \multicolumn{1}{c|}{$w_1=+1$}                                                               & $W=+1$                              \\ \cline{4-6} 
\multicolumn{1}{|c|}{}                       & \multicolumn{1}{c|}{}                       & \multicolumn{1}{c|}{}                        & \multicolumn{1}{c|}{15}                      & \multicolumn{1}{c|}{\begin{tabular}[c]{@{}c@{}}$w_1=+1$\\ $w_2=-1$\\ $w_3=+1$\end{tabular}} & $W=+1$                              \\ \cline{4-6} 
\multicolumn{1}{|c|}{}                       & \multicolumn{1}{c|}{}                       & \multicolumn{1}{c|}{}                        & \multicolumn{1}{c|}{20}                      & \multicolumn{1}{c|}{$w_1=+1$}                                                               & $W=+1$                              \\ \hline
\multicolumn{1}{|c|}{\multirow{4}{*}{0.005}} & \multicolumn{1}{c|}{\multirow{4}{*}{0.234}} & \multicolumn{1}{c|}{\multirow{4}{*}{-0.414}} & \multicolumn{1}{c|}{4}                       & \multicolumn{1}{c|}{$w_1=+1$}                                                               & $W=+1$                              \\ \cline{4-6} 
\multicolumn{1}{|c|}{}                       & \multicolumn{1}{c|}{}                       & \multicolumn{1}{c|}{}                        & \multicolumn{1}{c|}{10}                      & \multicolumn{1}{c|}{\begin{tabular}[c]{@{}c@{}}$w_1=+1$\\ $w_2=-1$\\ $w_3=+1$\end{tabular}} & $W=+1$                              \\ \cline{4-6} 
\multicolumn{1}{|c|}{}                       & \multicolumn{1}{c|}{}                       & \multicolumn{1}{c|}{}                        & \multicolumn{1}{c|}{\multirow{2}{*}{20}}     & \multicolumn{1}{c|}{\multirow{2}{*}{$w_1=+1$}}                                              & \multirow{2}{*}{$W=+1$}             \\
\multicolumn{1}{|c|}{}                       & \multicolumn{1}{c|}{}                       & \multicolumn{1}{c|}{}                        & \multicolumn{1}{c|}{}                        & \multicolumn{1}{c|}{}                                                                       &                                     \\ \hline
\multicolumn{1}{|c|}{\multirow{3}{*}{0.005}} & \multicolumn{1}{c|}{\multirow{3}{*}{0.034}} & \multicolumn{1}{c|}{\multirow{3}{*}{-0.114}} & \multicolumn{1}{c|}{1}                       & \multicolumn{1}{c|}{$w_1=+1$}                                                               & $W=+1$                              \\ \cline{4-6} 
\multicolumn{1}{|c|}{}                       & \multicolumn{1}{c|}{}                       & \multicolumn{1}{c|}{}                        & \multicolumn{1}{c|}{10}                      & \multicolumn{1}{c|}{\begin{tabular}[c]{@{}c@{}}$w_1=+1$\\ $w_2=-1$\\ $w_3=+1$\end{tabular}} & $W=+1$                              \\ \cline{4-6} 
\multicolumn{1}{|c|}{}                       & \multicolumn{1}{c|}{}                       & \multicolumn{1}{c|}{}                        & \multicolumn{1}{c|}{20}                      & \multicolumn{1}{c|}{$w_1=+1$}                                                               & $W=+1$                              \\ \hline
\multicolumn{1}{|c|}{\multirow{4}{*}{0.1}}   & \multicolumn{1}{c|}{\multirow{4}{*}{0.234}} & \multicolumn{1}{c|}{\multirow{4}{*}{-0.114}} & \multicolumn{1}{c|}{5}                       & \multicolumn{1}{c|}{$w_1=+1$}                                                               & $W=+1$                              \\ \cline{4-6} 
\multicolumn{1}{|c|}{}                       & \multicolumn{1}{c|}{}                       & \multicolumn{1}{c|}{}                        & \multicolumn{1}{c|}{10}                      & \multicolumn{1}{c|}{$w_1=+1$}                                                               & $W=+1$                              \\ \cline{4-6} 
\multicolumn{1}{|c|}{}                       & \multicolumn{1}{c|}{}                       & \multicolumn{1}{c|}{}                        & \multicolumn{1}{c|}{\multirow{2}{*}{15}}     & \multicolumn{1}{c|}{\multirow{2}{*}{$w_1=+1$}}                                              & \multirow{2}{*}{$W=+1$}             \\
\multicolumn{1}{|c|}{}                       & \multicolumn{1}{c|}{}                       & \multicolumn{1}{c|}{}                        & \multicolumn{1}{c|}{}                        & \multicolumn{1}{c|}{}                                                                       &                                     \\ \hline
\end{tabular}
\caption{Winding numbers and Topological numbers for higher order QED corrected EHAdS black hole in canonical ensemble for negative $a$}
\label{HOEHBH_Negative_a}
\end{table}

\subsection{For positive $a$}
\begin{figure}[ht]
	\centering
		\begin{subfigure}{0.5\textwidth}
			\centering			\includegraphics[width=0.9\linewidth]{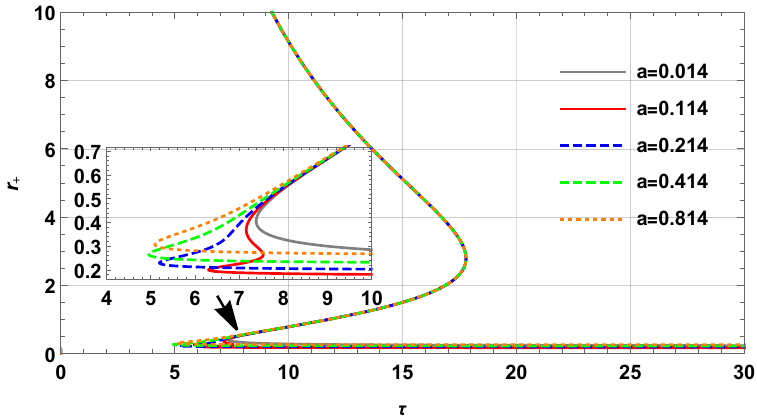}
			\caption{Zero points of $\phi$ for different positive values of $a$. Here $\beta=0.11$, $P=0.005$ and $Q=0.234$.}		\label{Fig:HOEHBH_Positive_a_(a_Variation)}
		\end{subfigure}%
		\begin{subfigure}{0.5\textwidth}
			\centering
		\includegraphics[width=0.9\linewidth]{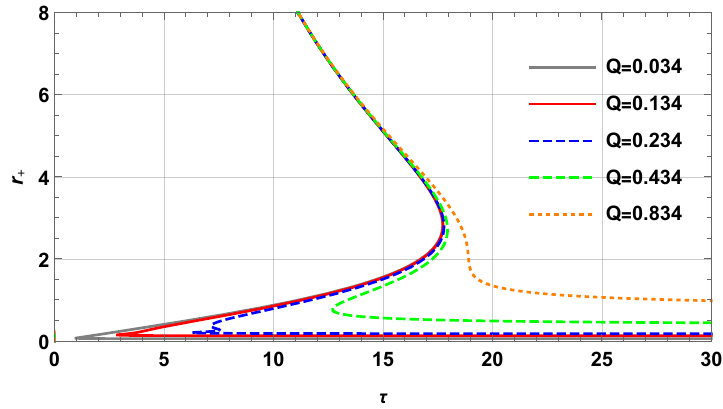}
			\caption{Zero points of $\phi$ for different values of $Q$. Here $\beta=0.11$, $a=0.114$ and $P=0.005$}			\label{Fig:HOEHBH_Positive_a_(Q_Variation)}
		\end{subfigure} \\
		\begin{subfigure}{0.5\textwidth}
			\centering			\includegraphics[width=0.9\linewidth]{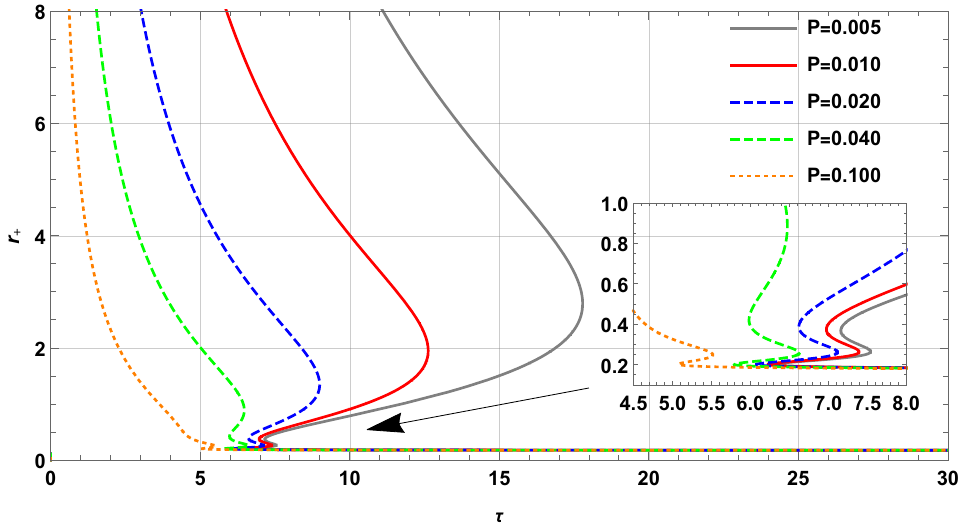}
			\caption{Zero points of $\phi$ for different values of $P$. Here $\beta=0.11$, $a=0.114$ and $Q=0.36$}			\label{Fig:HOEHBH_Positive_a_(P_Variation)}
		\end{subfigure}%
	\caption{$\tau$ vs $r_+$ plot with different parameter variations for higher-order QED corrected EHAdS black hole in canonical ensemble (for positive $a$).}
	\label{Fig:HOEHBH_Tau_Plot_All_Positive}
	\end{figure}

\begin{figure}[ht]
	\centering
		\begin{subfigure}{0.5\textwidth}
			\centering
			\includegraphics[width=0.9\linewidth]{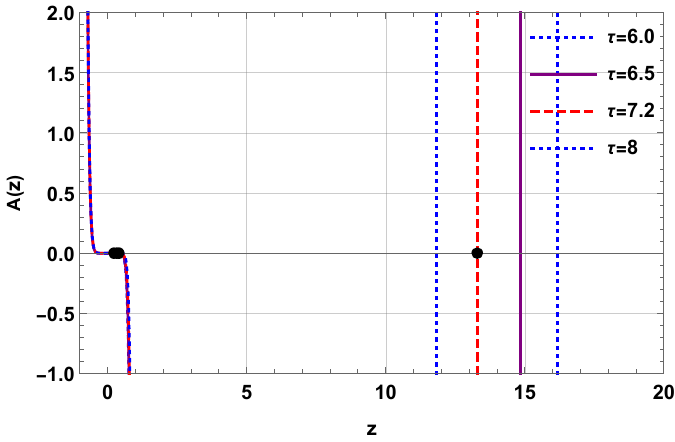}
			\caption{All roots of the polynomial $\mathcal{A}(z)$ for $\tau=7.2$}
			\label{Fig:Polynomial_Plot_HO_a_Positive_0dot114_P_0dot005_Q_0dot234_beta_0dot11}
		\end{subfigure}%
		\begin{subfigure}{0.5\textwidth}
			\centering
			\includegraphics[width=0.9\linewidth]{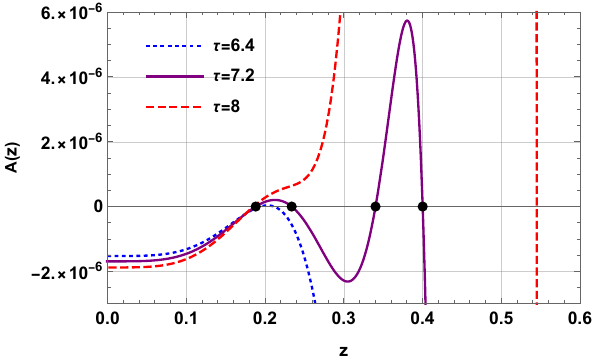}
			\caption{Enlarged view of the two smaller positive roots of $\mathcal{A}(z)$}
			\label{Fig:Polynomial_Plot_HO_a_Positive_0dot114_P_0dot005_Q_0dot234_beta_0dot11_Small_Region}
		\end{subfigure} \\
	\caption{Plot of the polynomial function $\mathcal{A}(z)$ for $\tau=6,~6.5,~7.2,\text{and}~8$. The black dots represents the roots for $\tau=7.2$.}
	\label{Fig:HOEHBH_Poly_Plot_All_Positive}
	\end{figure}

We repeat our calculation of a positive value of EH parameter $a$. The zero points of $\phi$ i.e. the variation of $\tau(r_+)$ curve for $\beta=0.11$ and different values of $P$, $Q$, and $a$ are shown in \autoref{Fig:HOEHBH_Tau_Plot_All_Positive}. In this case, we observe either $3$ or $5$ different branches of $\tau(r_+)$. For example, the red line in  \autoref{Fig:HOEHBH_Positive_a_(a_Variation)}, which is for $(P,Q,a,\beta)=(0.005,0.234,0.114,0.11)$ shows five different branches of $\tau(r_+)$ curve. In this parameter combination the polynomial function is given as 
\begin{equation}
\mathcal{A}(z)=-2.346915927219771\times 10^{-7} \tau +0.502655 \tau  z^{12}+4 \tau  z^{10}-0.219024 \tau  z^8+0.000341797 \tau  z^4-16 \pi  z^{11}
\end{equation}
The roots of $\mathcal{A}(z)$ for different values of $\tau$ is shown in \autoref{Fig:HOEHBH_Poly_Plot_All_Positive}. For $\tau=7.2$ we have five real positive roots of $\mathcal{A}(z)$ which are $z_1=0.188142$, $z_2=0.233831$, $z_3=0.340349$, $z_4=0.400221$ and $z_5=13.2903$. From the residue theorem we find that the corresponding winding numbers are $w_1=+1$, $w_2=-1$, $w_3=+1$, $w_4=-1$, and $w_5=+1$. The topological number is hence $W=w_1+w_2+w_3+w_4+w_5=+1$. The winding numbers $w_1=w_3=w_5=+1$ correspond to the stable black hole region whereas the winding numbers $w_2=w_4=-1$ correspond to unstable black hole region. The winding numbers as well as the topological numbers with different values of parameters are shown in the \autoref{HOEHBH_Positive_a}.

\begin{table}[ht]
\centering
\begin{tabular}{|cccccc|}
\hline
\multicolumn{6}{|c|}{Higher order EHAdS black hole ($\beta = 0.11$)}                                                                                                                                                                                                                                                                                     \\ \hline
\multicolumn{3}{|c|}{Parameters}                                                                                                          & \multicolumn{1}{c|}{\multirow{2}{*}{$\tau$}} & \multicolumn{1}{c|}{\multirow{2}{*}{Winding number}}                                                                 & \multirow{2}{*}{Topological number} \\ \cline{1-3}
\multicolumn{1}{|c|}{P}                      & \multicolumn{1}{c|}{Q}                      & \multicolumn{1}{c|}{a}                       & \multicolumn{1}{c|}{}                        & \multicolumn{1}{c|}{}                                                                                                &                                     \\ \hline
\multicolumn{1}{|c|}{\multirow{4}{*}{0.005}} & \multicolumn{1}{c|}{\multirow{4}{*}{0.234}} & \multicolumn{1}{c|}{\multirow{4}{*}{0.114}}  & \multicolumn{1}{c|}{6}                       & \multicolumn{1}{c|}{$w_1=+1$}                                                                                        & $W=+1$                              \\ \cline{4-6} 
\multicolumn{1}{|c|}{}                       & \multicolumn{1}{c|}{}                       & \multicolumn{1}{c|}{}                        & \multicolumn{1}{c|}{6.5}                     & \multicolumn{1}{c|}{\begin{tabular}[c]{@{}c@{}}$w_1=+1$\\ $w_2=-1$\\ $w_3=+1$\end{tabular}}                          & $W=+1$                              \\ \cline{4-6} 
\multicolumn{1}{|c|}{}                       & \multicolumn{1}{c|}{}                       & \multicolumn{1}{c|}{}                        & \multicolumn{1}{c|}{7.2}                     & \multicolumn{1}{c|}{\begin{tabular}[c]{@{}c@{}}$w_1=+1$\\ $w_2=-1$\\ $w_3=+1$\\ $w_4=-1$\\ \\ $w_5=+1$\end{tabular}} & $W=+1$                              \\ \cline{4-6} 
\multicolumn{1}{|c|}{}                       & \multicolumn{1}{c|}{}                       & \multicolumn{1}{c|}{}                        & \multicolumn{1}{c|}{8}                       & \multicolumn{1}{c|}{$w_1=+1$}                                                                                        & $W=+1$                              \\ \hline
\multicolumn{1}{|c|}{\multirow{4}{*}{0.005}} & \multicolumn{1}{c|}{\multirow{4}{*}{0.234}} & \multicolumn{1}{c|}{\multirow{4}{*}{0.414}}  & \multicolumn{1}{c|}{4}                       & \multicolumn{1}{c|}{$w_1=+1$}                                                                                        & $W=+1$                              \\ \cline{4-6} 
\multicolumn{1}{|c|}{}                       & \multicolumn{1}{c|}{}                       & \multicolumn{1}{c|}{}                        & \multicolumn{1}{c|}{10}                      & \multicolumn{1}{c|}{\begin{tabular}[c]{@{}c@{}}$w_1=+1$\\ $w_2=-1$\\ $w_3=+1$\end{tabular}}                          & $W=+1$                              \\ \cline{4-6} 
\multicolumn{1}{|c|}{}                       & \multicolumn{1}{c|}{}                       & \multicolumn{1}{c|}{}                        & \multicolumn{1}{c|}{\multirow{2}{*}{20}}     & \multicolumn{1}{c|}{\multirow{2}{*}{$w_1=+1$}}                                                                       & \multirow{2}{*}{$W=+1$}             \\
\multicolumn{1}{|c|}{}                       & \multicolumn{1}{c|}{}                       & \multicolumn{1}{c|}{}                        & \multicolumn{1}{c|}{}                        & \multicolumn{1}{c|}{}                                                                                                &                                     \\ \hline
\multicolumn{1}{|c|}{\multirow{3}{*}{0.005}} & \multicolumn{1}{c|}{\multirow{3}{*}{0.034}} & \multicolumn{1}{c|}{\multirow{3}{*}{0.114}}  & \multicolumn{1}{c|}{2}                       & \multicolumn{1}{c|}{\begin{tabular}[c]{@{}c@{}}$w_1=+1$\\ $w_2=-1$\\ $w_3=+1$\end{tabular}}                          & $W=+1$                              \\ \cline{4-6} 
\multicolumn{1}{|c|}{}                       & \multicolumn{1}{c|}{}                       & \multicolumn{1}{c|}{}                        & \multicolumn{1}{c|}{10}                      & \multicolumn{1}{c|}{\begin{tabular}[c]{@{}c@{}}$w_1=+1$\\ $w_2=-1$\\ $w_3=+1$\end{tabular}}                          & $W=+1$                              \\ \cline{4-6} 
\multicolumn{1}{|c|}{}                       & \multicolumn{1}{c|}{}                       & \multicolumn{1}{c|}{}                        & \multicolumn{1}{c|}{20}                      & \multicolumn{1}{c|}{$w_1=+1$}                                                                                        & $W=+1$                              \\ \hline
\multicolumn{1}{|c|}{\multirow{4}{*}{0.1}}   & \multicolumn{1}{c|}{\multirow{4}{*}{0.234}} & \multicolumn{1}{c|}{\multirow{4}{*}{0.114}} & \multicolumn{1}{c|}{5}                       & \multicolumn{1}{c|}{$w_1=+1$}                                                                                        & $W=+1$                              \\ \cline{4-6} 
\multicolumn{1}{|c|}{}                       & \multicolumn{1}{c|}{}                       & \multicolumn{1}{c|}{}                        & \multicolumn{1}{c|}{5.2}                     & \multicolumn{1}{c|}{\begin{tabular}[c]{@{}c@{}}$w_1=+1$\\ $w_2=-1$\\ $w_3=+1$\end{tabular}}                          & $W=+1$                              \\ \cline{4-6} 
\multicolumn{1}{|c|}{}                       & \multicolumn{1}{c|}{}                       & \multicolumn{1}{c|}{}                        & \multicolumn{1}{c|}{\multirow{2}{*}{6}}      & \multicolumn{1}{c|}{\multirow{2}{*}{$w_1=+1$}}                                                                       & \multirow{2}{*}{$W=+1$}             \\
\multicolumn{1}{|c|}{}                       & \multicolumn{1}{c|}{}                       & \multicolumn{1}{c|}{}                        & \multicolumn{1}{c|}{}                        & \multicolumn{1}{c|}{}                                                                                                &                                     \\ \hline
\end{tabular}
\caption{Winding numbers and Topological numbers for higher order QED corrected EHAdS black hole in canonical ensemble for positive $a$}
\label{HOEHBH_Positive_a}
\end{table}

\section{4D Euler-Heisenberg AdS black hole in grand canonical ensemble}
\label{The Euler-Heisenberg-AdS black hole in grand canonical ensemble}
The thermodynamics of black holes may be different in different ensembles. Therefore, we extend our study to the grand canonical ensemble.
In this ensemble, the charge $Q$ is written in terms of the electric potential $\phi$ as
\begin{equation}
Q=\frac{-\sqrt[3]{15} \Bigl\{r_+^5 \left(\sqrt{3} \sqrt{a^3 \left(27 a \phi_e ^2-40 r_+^2\right)}+9 a^2 \phi_e \right)\Bigr\}^{2/3}-2\times 15^{2/3} a r_+^4}{3 a \sqrt[3]{r_+^5 \left(\sqrt{3} \sqrt{a^3 \left(27 a \phi_e ^2-40 r_+^2\right)}+9 a^2 \phi_e \right)}}
\end{equation}
and mass is written as 
\begin{equation}
\begin{aligned}
\mathcal{M} =& M-Q\phi_e \\
		   =&\frac{3 \sqrt[6]{3} 5^{2/3} r_+^9 \phi_e  \left(\sqrt{a^3 \left(27 a \phi_e ^2-40 r_+^2\right)}+3 \sqrt{3} a^2 \phi_e \right)}{A^{4/3}}-\frac{10\times 5^{2/3} a r_+^{11}}{\sqrt[3]{3} A^{4/3}}+\frac{\sqrt[3]{15} \sqrt[3]{A} \phi_e }{4 a} \\
		   &+\frac{r_+^3 (4 \pi  a P+5)}{3 a}+\frac{5 \sqrt[3]{5} r_+^7}{3^{2/3} A^{2/3}}+\frac{r_+}{2},
\end{aligned}
\end{equation}
where $A=r_+^5 \Bigl\{\sqrt{3} \sqrt{a^3 \left(27 a \phi_e ^2-40 r_+^2\right)}+9 a^2 \phi_e \Bigr\}$ and $M$ is given by \eqref{Eq:EHBHMass}. 
The free energy is modified as 
\begin{equation}
\label{Eq:EHBH_GrandCan_FreeEnergy}
\mathcal{F}=\mathcal{M}-\frac{S}{\tau }.
\end{equation}
The expression for $\tau$ can be calculated by $\partial_{r_+}\mathcal{F}=0$.  
From \eqref{Eq:GeneralComplexFunction} we calculate the complex function $\mathcal{R}(z)$, the denominator of which is the polynomial function $\mathcal{A}(z)$ (see \autoref{EHBH_Grand_Can_ComF} and \autoref{EHBH_Grand_Can_PolyF}).

\subsection{For negative $a$}
\begin{figure}[ht]
	\centering
		\begin{subfigure}{0.5\textwidth}
			\centering
			\includegraphics[width=0.9\linewidth]{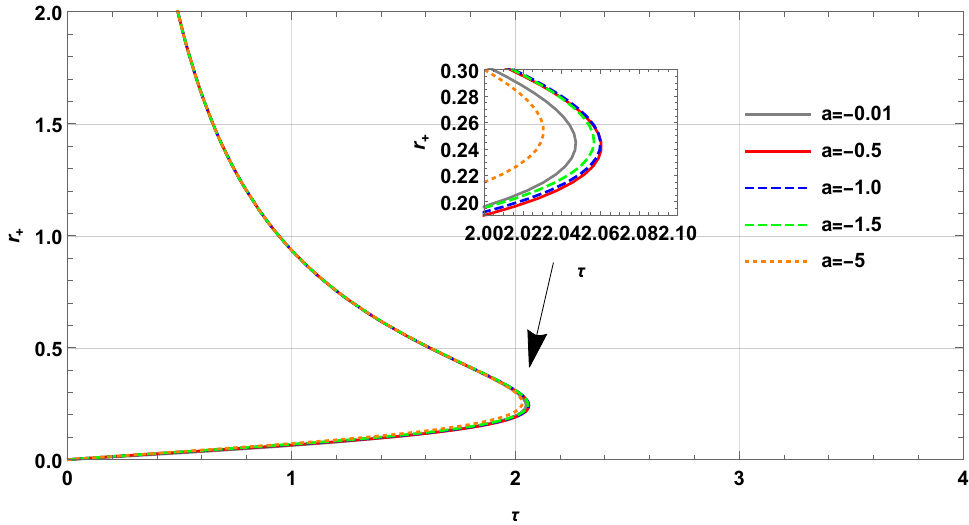}
			\caption{Zero points of $\phi$ for different negative values of $a$. Here $P=0.5$ and $\phi_e=0.5$.}
			\label{Fig:EHBH_GrandCan_Negative_a_a_Variation}
		\end{subfigure}%
		\begin{subfigure}{0.5\textwidth}
			\centering
			\includegraphics[width=0.9\linewidth]{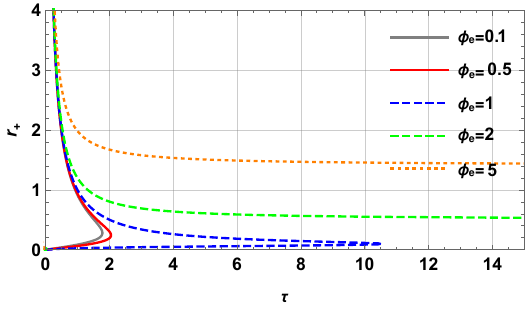}
			\caption{Zero points of $\phi$ for different values of $\phi_e$. Here $a=-0.1$ and $P=0.5$}
			\label{Fig:EHBH_GrandCan_Negative_a_phi_Variation}
		\end{subfigure} \\
		\begin{subfigure}{0.5\textwidth}
			\centering
			\includegraphics[width=0.9\linewidth]{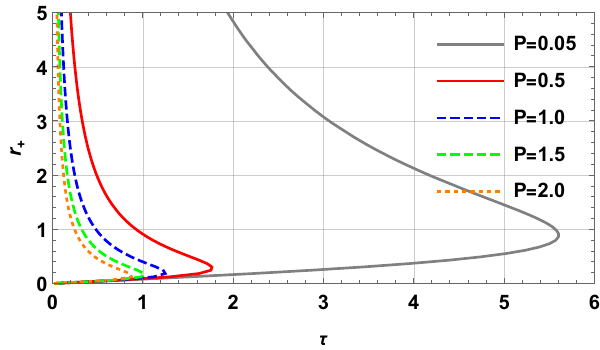}
			\caption{Zero points of $\phi_e$ for different values of $P$. Here $a=-0.1$ and $\phi_e=0.01$}
			\label{Fig:EHBH_GrandCan_Negative_a_P_Variation}
		\end{subfigure}%
	\caption{$\tau$ vs $r_+$ plot with different parameter variations for EHAdS black hole in grand canonical ensemble (for negative $a$).}
	\label{Fig:EHBH_GrandCan_Tau_Plot_All_Negative}
	\end{figure}

\begin{figure}[h!]
	\centerline{
	\includegraphics[scale=0.4]{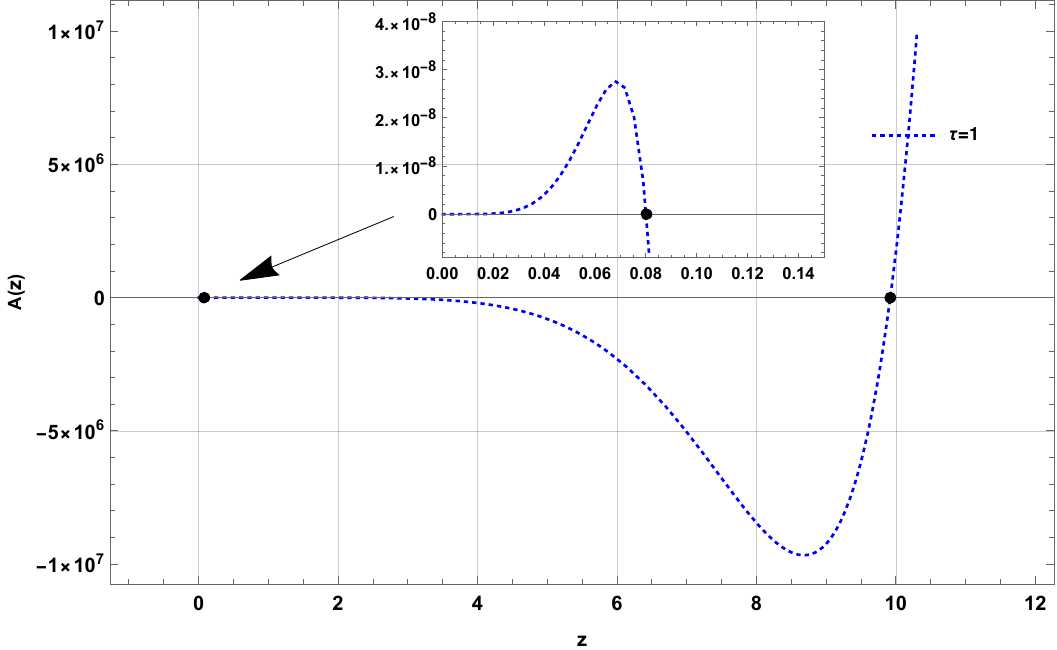}}
	\caption	{Plot of the polynomial function $\mathcal{A}(z)$ for $\tau=1$. The black dots represent the roots of the polynomial.}	\label{Fig:Poly_Plot_EHBH_a_Neg_a_0dot1_P_0dot05_phi_0dot01}
	\end{figure}

The $\tau(r_+)$ curve for parameters is shown in \autoref{Fig:EHBH_GrandCan_Tau_Plot_All_Negative}. Here we see that for different sets of parameters we have two branches of the $\tau$ curve. In \autoref{Fig:EHBH_GrandCan_Negative_a_a_Variation} we see that there is slight variation of the $\tau$ curve with the variation of EH parameter $a$. For higher values of $\phi_e$ (such as $\phi_e=2,~5$) in \autoref{Fig:EHBH_GrandCan_Negative_a_phi_Variation} the lower part is not visible but they follow the similar pattern as the other curves. We plot polynomial function $\mathcal{A}(z)$ for a parameter set $(P,\phi_e,a)=(0.05,0.01,-0.1)$ in \autoref{Fig:Poly_Plot_EHBH_a_Neg_a_0dot1_P_0dot05_phi_0dot01} and the roots are found to be at $z_1=9.91978707$ and $z_2=0.0802129245$. Here we choose $\tau=1$ Interestingly, we have found that the roots indicate a second order pole of the complex function $\mathcal{R}(z)$. Using residue method we calculate the winding number. Corresponding to $z_1$ and $z_2$ the winding numbers are respectively $w_1=+1$ and $w_2=-1$ yielding topological number $W=w_1+w_2=0$. 
We have calculated the winding numbers and topological numbers for different sets of parameters which is represented in the \autoref{EHBH_Grand_Can_Neg_a}. 
\begin{table}[]
\begin{tabular}{|cccccc|}
\hline
\multicolumn{6}{|c|}{EH black hole in grand canonical ensemble}                                                                                                                                                                                              \\ \hline
\multicolumn{3}{|c|}{Parameters}                                                     & \multicolumn{1}{c|}{\multirow{2}{*}{$\tau$}} & \multicolumn{1}{c|}{\multirow{2}{*}{Winding number}}                             & \multirow{2}{*}{Topological number} \\ \cline{1-3}
\multicolumn{1}{|c|}{P}    & \multicolumn{1}{c|}{$\phi$} & \multicolumn{1}{c|}{a}    & \multicolumn{1}{c|}{}                        & \multicolumn{1}{c|}{}                                                            &                                     \\ \hline
\multicolumn{1}{|c|}{0.05} & \multicolumn{1}{c|}{0.01}   & \multicolumn{1}{c|}{-0.1} & \multicolumn{1}{c|}{1}                       & \multicolumn{1}{c|}{\begin{tabular}[c]{@{}c@{}}$w_1=+1$\\ $w_2=-1$\end{tabular}} & $W=0$                               \\ \hline
\multicolumn{1}{|c|}{0.05} & \multicolumn{1}{c|}{0.01}   & \multicolumn{1}{c|}{-1}   & \multicolumn{1}{c|}{1}                       & \multicolumn{1}{c|}{\begin{tabular}[c]{@{}c@{}}$w_1=+1$\\ $w_2=-1$\end{tabular}} & $W=0$                               \\ \hline
\multicolumn{1}{|c|}{0.05} & \multicolumn{1}{c|}{0.01}   & \multicolumn{1}{c|}{-5}   & \multicolumn{1}{c|}{1}                       & \multicolumn{1}{c|}{\begin{tabular}[c]{@{}c@{}}$w_1=+1$\\ $w_2=-1$\end{tabular}} & $W=0$                               \\ \hline
\multicolumn{1}{|c|}{0.5}  & \multicolumn{1}{c|}{0.01}   & \multicolumn{1}{c|}{-0.1} & \multicolumn{1}{c|}{1}                       & \multicolumn{1}{c|}{\begin{tabular}[c]{@{}c@{}}$w_1=+1$\\ $w_2=-1$\end{tabular}} & $W=0$                               \\ \hline
\multicolumn{1}{|c|}{1.5}  & \multicolumn{1}{c|}{0.01}   & \multicolumn{1}{c|}{-0.1} & \multicolumn{1}{c|}{0.5}                     & \multicolumn{1}{c|}{\begin{tabular}[c]{@{}c@{}}$w_1=+1$\\ $w_2=-1$\end{tabular}} & $W=0$                               \\ \hline
\multicolumn{1}{|c|}{0.5}  & \multicolumn{1}{c|}{0.5}    & \multicolumn{1}{c|}{-0.1} & \multicolumn{1}{c|}{1}                       & \multicolumn{1}{c|}{\begin{tabular}[c]{@{}c@{}}$w_1=+1$\\ $w_2=-1$\end{tabular}} & $W=0$                               \\ \hline
\multicolumn{1}{|c|}{0.5}  & \multicolumn{1}{c|}{2}      & \multicolumn{1}{c|}{-0.1} & \multicolumn{1}{c|}{1}                       & \multicolumn{1}{c|}{\begin{tabular}[c]{@{}c@{}}$w_1=+1$\\ $w_2=-1$\end{tabular}} & $W=0$                               \\ \hline
\end{tabular}
\caption{Winding numbers and topological numbers for EHAdS black hole in grand canonical ensemble for negative $a$}
\label{EHBH_Grand_Can_Neg_a}
\end{table}

\subsection{Positive $a$}
\begin{figure}[ht]
	\centering
		\begin{subfigure}{0.5\textwidth}
			\centering
			\includegraphics[width=0.9\linewidth]{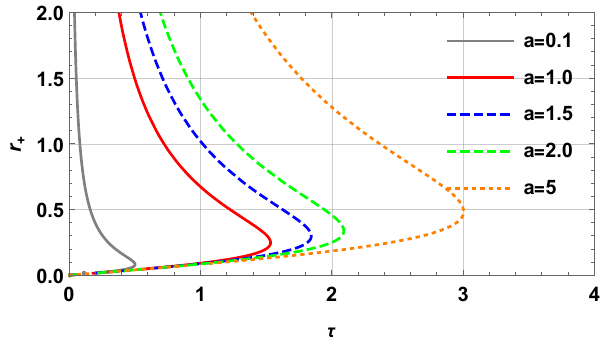}
			\caption{Zero points of $\phi$ for different negative values of $a$. Here $P=0.5$ and $\phi_e=0.5$.}
			\label{Fig:EHBH_GrandCan_Positive_a_a_Variation}
		\end{subfigure}%
		\begin{subfigure}{0.5\textwidth}
			\centering
			\includegraphics[width=0.9\linewidth]{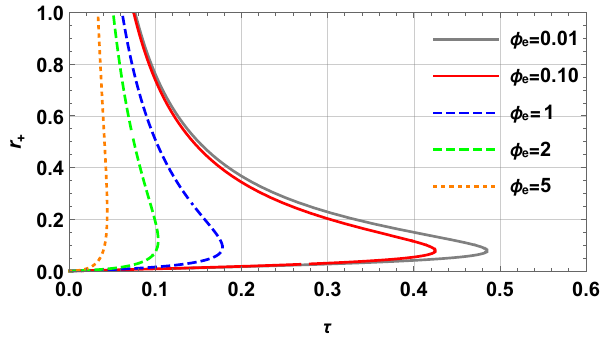}
			\caption{Zero points of $\phi$ for different values of $\phi_e$. Here $a=0.1$ and $P=0.5$}
			\label{Fig:EHBH_GrandCan_Positive_a_phi_Variation}
		\end{subfigure} \\
		\begin{subfigure}{0.5\textwidth}
			\centering
			\includegraphics[width=0.9\linewidth]{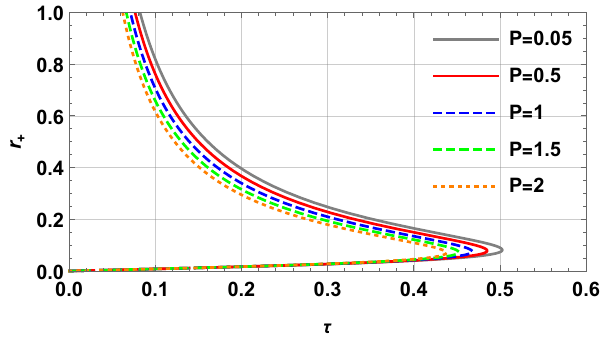}
			\caption{Zero points of $\phi_e$ for different values of $P$. Here $a=0.1$ and $\phi_e=0.01$}
			\label{Fig:EHBH_GrandCan_Positive_a_P_Variation}
		\end{subfigure}%
	\caption{$\tau$ vs $r_+$ plot with different parameter variations for EHAdS black hole in grand canonical ensemble (for positive $a$).}
	\label{Fig:EHBH_GrandCan_Tau_Plot_All_Positive}
	\end{figure}

\begin{figure}[h!]
	\centerline{
	\includegraphics[scale=0.55]{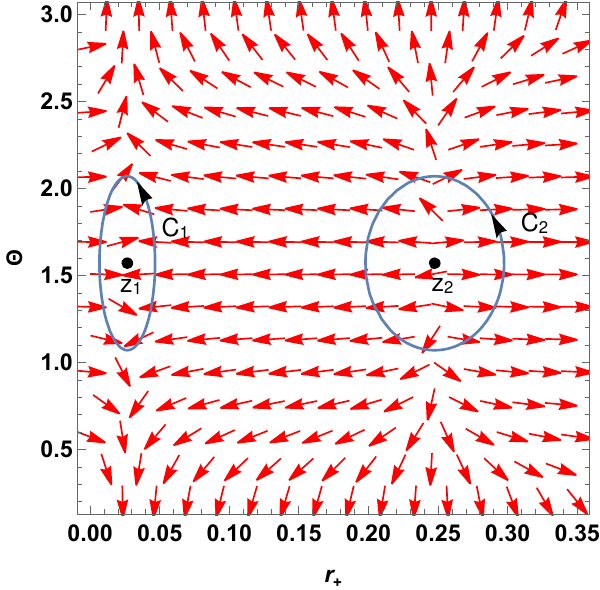}}
	\caption	{Vector plot of $n$. The black dots represent the zero points.}	\label{Fig:EHBH_GrandCan_a_Posi_Vector_Plot}
	\end{figure}

Now, we change the sign of the EH parameter $a$ to positive and study the thermodynamic topology. For simplicity, in this case we calculate the winding numbers by drawing contours around the zero points. Using \eqref{Eq:EHBH_GrandCan_FreeEnergy} and \eqref{Eq:Vector_Field} we calculate the vector field $\phi$ and then, normalized vector field 
as
\begin{equation}
n=\Bigl( \frac{\phi_{r_+}}{|\phi|},\frac{\phi_{\Theta}}{|\phi|} \Bigr)\quad \quad \text{where,} \quad \phi_{r_+}=\frac{\partial\mathcal{F}}{\partial r_+}, ~ \phi_{\Theta}= -\cot\Theta\csc\Theta.
\end{equation}
The zero point can be calculated by setting $\Theta=\pi/2$ in $\phi_{r_+}$. The plot of $\tau(r_+)$ curve is shown in \autoref{Fig:EHBH_GrandCan_Tau_Plot_All_Positive} where we observe two black hole branches. 
For $(P,\phi_e,a)=(0.05,0.01,0.1)$ and $\tau=0.3$ the vector plot of $n$ is shown in \autoref{Fig:EHBH_GrandCan_a_Posi_Vector_Plot}. The zero points are situated at $(\Theta,r_{+})=(\pi/2,0.24758) ~ \text{and}  ~ (\pi/2,0.0267)$. Two contours $C_1$ and $C_2$ (see \autoref{Fig:EHBH_GrandCan_a_Posi_Vector_Plot}) are drawn enclosing the zero points which are defined as \eqref{Contour}. The zero points $z_1=0.24758$ and $z_2=0.0267$ respectively corresponds to $w_1=+1$ and $w_2=-1$ winding number. The topological number is hence $W=+1-1=0$. The winding numbers and the corresponding topological number for various set of parameters are shown in \autoref{EHBH_Grand_Can_Posi_a}.

\begin{table}[]
\begin{tabular}{|cccccc|}
\hline
\multicolumn{6}{|c|}{EH black hole}                                                                                                                                                                                                                         \\ \hline
\multicolumn{3}{|c|}{Parameters}                                                    & \multicolumn{1}{c|}{\multirow{2}{*}{$\tau$}} & \multicolumn{1}{c|}{\multirow{2}{*}{Winding number}}                             & \multirow{2}{*}{Topological number} \\ \cline{1-3}
\multicolumn{1}{|c|}{P}    & \multicolumn{1}{c|}{$\phi$} & \multicolumn{1}{c|}{a}   & \multicolumn{1}{c|}{}                        & \multicolumn{1}{c|}{}                                                            &                                     \\ \hline
\multicolumn{1}{|c|}{0.05} & \multicolumn{1}{c|}{0.01}   & \multicolumn{1}{c|}{0.1} & \multicolumn{1}{c|}{0.3}                     & \multicolumn{1}{c|}{\begin{tabular}[c]{@{}c@{}}$w_1=+1$\\ $w_2=-1$\end{tabular}} & $W=0$                               \\ \hline
\multicolumn{1}{|c|}{0.05} & \multicolumn{1}{c|}{0.01}   & \multicolumn{1}{c|}{1}   & \multicolumn{1}{c|}{1}                       & \multicolumn{1}{c|}{\begin{tabular}[c]{@{}c@{}}$w_1=+1$\\ $w_2=-1$\end{tabular}} & $W=0$                               \\ \hline
\multicolumn{1}{|c|}{0.05} & \multicolumn{1}{c|}{0.01}   & \multicolumn{1}{c|}{2}   & \multicolumn{1}{c|}{1}                       & \multicolumn{1}{c|}{\begin{tabular}[c]{@{}c@{}}$w_1=+1$\\ $w_2=-1$\end{tabular}} & $W=0$                               \\ \hline
\multicolumn{1}{|c|}{0.5}  & \multicolumn{1}{c|}{0.01}   & \multicolumn{1}{c|}{0.1} & \multicolumn{1}{c|}{0.3}                     & \multicolumn{1}{c|}{\begin{tabular}[c]{@{}c@{}}$w_1=+1$\\ $w_2=-1$\end{tabular}} & $W=0$                               \\ \hline
\multicolumn{1}{|c|}{1.5}  & \multicolumn{1}{c|}{0.01}   & \multicolumn{1}{c|}{0.1} & \multicolumn{1}{c|}{0.3}                     & \multicolumn{1}{c|}{\begin{tabular}[c]{@{}c@{}}$w_1=+1$\\ $w_2=-1$\end{tabular}} & $W=0$                               \\ \hline
\multicolumn{1}{|c|}{0.5}  & \multicolumn{1}{c|}{0.1}    & \multicolumn{1}{c|}{0.1} & \multicolumn{1}{c|}{0.35}                    & \multicolumn{1}{c|}{\begin{tabular}[c]{@{}c@{}}$w_1=+1$\\ $w_2=-1$\end{tabular}} & $W=0$                               \\ \hline
\multicolumn{1}{|c|}{0.5}  & \multicolumn{1}{c|}{2}      & \multicolumn{1}{c|}{0.1} & \multicolumn{1}{c|}{0.08}                    & \multicolumn{1}{c|}{\begin{tabular}[c]{@{}c@{}}$w_1=+1$\\ $w_2=-1$\end{tabular}} & $W=0$                               \\ \hline
\end{tabular}
\caption{Winding numbers and topological numbers for EHAdS black hole in grand canonical ensemble for positive $a$}
\label{EHBH_Grand_Can_Posi_a}
\end{table}

\section{Higher order QED corrected Euler-Heisenberg-AdS black hole in grand canonical ensemble}
\label{The higher order Euler-Heisenberg-AdS black hole in grand canonical ensemble}
In grand canonical ensemble the charge $Q$ is decomposed to the electric potential $\phi_e$. The electric potential is the conjugate of charge. To calculate $\phi_e$ we partially differentiate the mass $M'$ (see \eqref{Eq:HOEHBH_Mass}) with respect to $Q$. This provides us
\begin{equation}
\label{Eq:HOEHBH_electric_potential}
\phi_e=\frac{a^2 \beta  Q^5}{12 r_+^9}+\frac{120 Q r_+^4-12 a Q^3}{120 r_+^5}.
\end{equation}
Writing the charge $Q$ in terms of $\phi_e$ using \eqref{Eq:HOEHBH_electric_potential}, new mass can be calculated using 
\begin{equation}
\begin{aligned}
\mathcal{M} =& M'-Q\phi_e,
\end{aligned}
\end{equation}
where $M'$ is given by \eqref{Eq:HOEHBH_Mass}. The free energy is calculated as 
\begin{equation}
\label{Eq:HOEHGCanFreeEnergy}
\mathcal{F}=\mathcal{M}-\frac{S}{\tau }.
\end{equation}
In residue method the complex function $\mathcal{R}(z)$ is defined as \eqref{Eq:GeneralComplexFunction} and the denominator of this function is the polynomial function $\mathcal{A}(z)$ (see \autoref{HOEHBH_Grand_Can_ComF} and \autoref{HOEHBH_Grand_Can_PolyF}). We choose $\beta=0.11$ for our study.

\subsection{Negative $a$}
\begin{figure}[ht]
	\centering
		\begin{subfigure}{0.5\textwidth}
			\centering
			\includegraphics[width=0.9\linewidth]{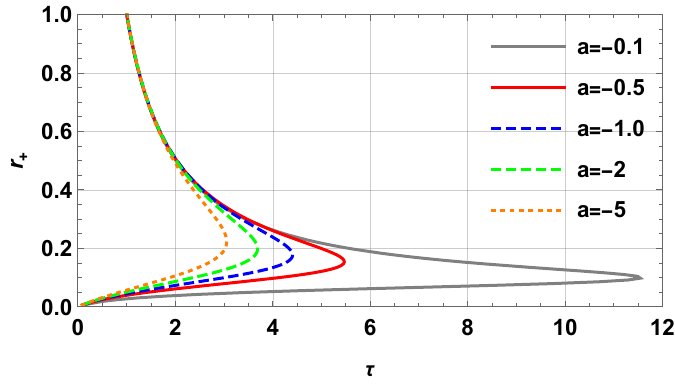}
			\caption{$\tau$ vs $r_{+}$ plot for different negative values of $a$. Here $P=0.5$ and $\phi_e=1$.}
			\label{Fig:HOEHBH_GrandCan_Negative_a_a_Variation}
		\end{subfigure}%
		\begin{subfigure}{0.5\textwidth}
			\centering
			\includegraphics[width=0.9\linewidth]{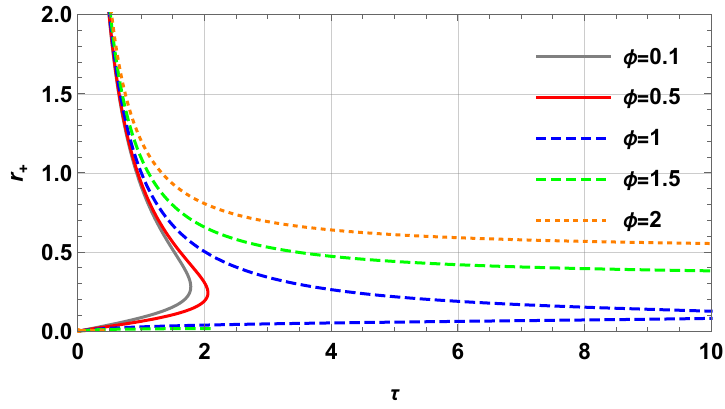}
			\caption{$\tau$ vs $r_{+}$ plot for different values of $\phi_e$. Here $a=-0.1$ and $P=0.5$}
			\label{Fig:HOEHBH_GrandCan_Negative_a_phi_Variation}
		\end{subfigure} \\
		\begin{subfigure}{0.5\textwidth}
			\centering
			\includegraphics[width=0.9\linewidth]{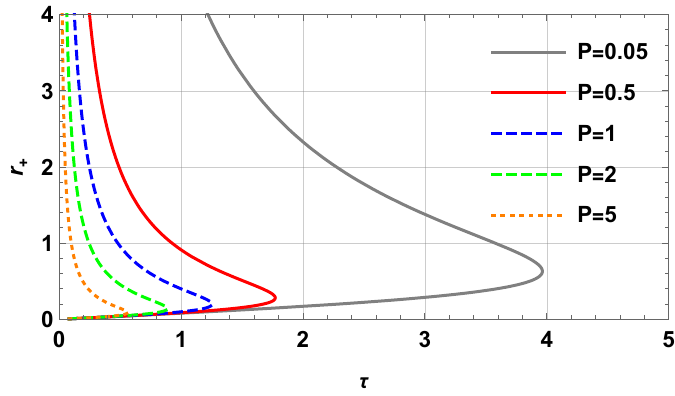}
			\caption{$\tau$ vs $r_{+}$ plot for different values of $P$. Here $a=-0.1$ and $\phi_e=0.01$}
			\label{Fig:HOEHBH_GrandCan_Negative_a_P_Variation}
		\end{subfigure}%
	\caption{$\tau$ vs $r_+$ plot with different parameter variations for higher order QED corrected EHAdS black hole in grand canonical ensemble (for negative $a$).}
	\label{Fig:HOEHBH_GrandCan_Tau_Plot_All_Negative}
	\end{figure}

\begin{figure}[h!]
	\centerline{
	\includegraphics[scale=0.6]{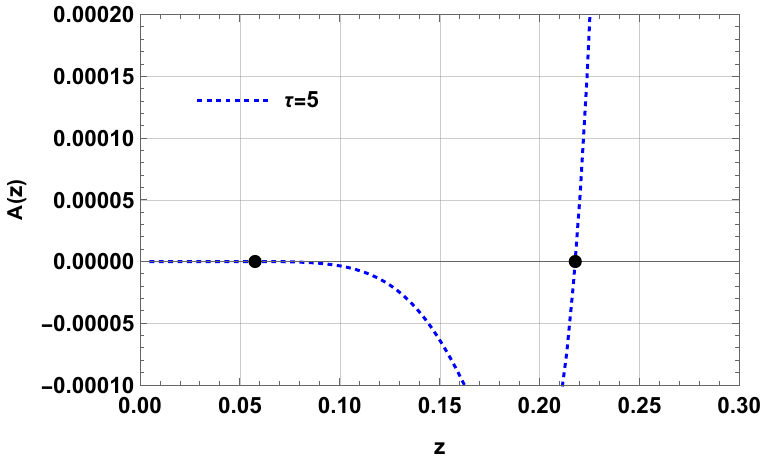}}
	\caption	{Plot of the polynomial function $\mathcal{A}(z)$ for $\tau=5$. The black dots represent the roots of the polynomial.}	\label{Fig:Poly_Plot_HOEHBH_a_Neg_a}
	\end{figure}

From \eqref{Eq:HOEHGCanFreeEnergy} we calculate the $\tau$ curve using $\frac{\partial \mathcal{F}}{\partial r_{+}}=0$. For different set of parameters the $\tau$ vs $r_{+}$ curve is shown in \autoref{Fig:HOEHBH_GrandCan_Tau_Plot_All_Negative}. For different combination of parameters we have two different branches of the $\tau(r_+)$ curve. To calculate the winding numbers we choose a set of parameters $(P,\phi_e,a,\beta)=(0.5,1,-0.1,0.11)$ and $\tau=5$. The polynomial function $\mathcal{A}(z)$ is shown in \autoref{Fig:Poly_Plot_HOEHBH_a_Neg_a}. The black dots represent the roots of $\mathcal{A}(z)$ which are $z_1=0.2178700113$ and $z_2=0.057573538$. From residue method the winding numbers corresponding to $z_1$ and $z_2$ are obtained as $w_1=+1$ and $w_2=-1$. The topological number is $W=+1-1=0$. For various set of parameters the winding number and the topological number are shown in \autoref{HOEHBH_Grand_Can_Neg_a}. 

\begin{table}[]
\begin{tabular}{|cccccc|}
\hline
\multicolumn{6}{|c|}{Higher order EHAdS black hole in grand canonical ensemble ($\beta = 0.11$)}                                                                                                                                                               \\ \hline
\multicolumn{3}{|c|}{Parameters}                                                    & \multicolumn{1}{c|}{\multirow{2}{*}{$\tau$}} & \multicolumn{1}{c|}{\multirow{2}{*}{Winding number}}                             & \multirow{2}{*}{Topological number} \\ \cline{1-3}
\multicolumn{1}{|c|}{P}   & \multicolumn{1}{c|}{$\phi$} & \multicolumn{1}{c|}{a}    & \multicolumn{1}{c|}{}                        & \multicolumn{1}{c|}{}                                                            &                                     \\ \hline
\multicolumn{1}{|c|}{0.5} & \multicolumn{1}{c|}{1}      & \multicolumn{1}{c|}{-0.1} & \multicolumn{1}{c|}{5}                       & \multicolumn{1}{c|}{\begin{tabular}[c]{@{}c@{}}$w_1=+1$\\ $w_2=-1$\end{tabular}} & $W=0$                               \\ \hline
\multicolumn{1}{|c|}{0.5} & \multicolumn{1}{c|}{1}      & \multicolumn{1}{c|}{-1}   & \multicolumn{1}{c|}{4}                       & \multicolumn{1}{c|}{\begin{tabular}[c]{@{}c@{}}$w_1=+1$\\ $w_2=-1$\end{tabular}} & $W=0$                               \\ \hline
\multicolumn{1}{|c|}{0.5} & \multicolumn{1}{c|}{1}      & \multicolumn{1}{c|}{-2}   & \multicolumn{1}{c|}{2}                       & \multicolumn{1}{c|}{\begin{tabular}[c]{@{}c@{}}$w_1=+1$\\ $w_2=-1$\end{tabular}} & $W=0$                               \\ \hline
\multicolumn{1}{|c|}{0.5} & \multicolumn{1}{c|}{0.1}    & \multicolumn{1}{c|}{-0.1} & \multicolumn{1}{c|}{1}                       & \multicolumn{1}{c|}{\begin{tabular}[c]{@{}c@{}}$w_1=+1$\\ $w_2=-1$\end{tabular}} & $W=0$                               \\ \hline
\multicolumn{1}{|c|}{0.5} & \multicolumn{1}{c|}{0.5}    & \multicolumn{1}{c|}{-0.1} & \multicolumn{1}{c|}{2}                       & \multicolumn{1}{c|}{\begin{tabular}[c]{@{}c@{}}$w_1=+1$\\ $w_2=-1$\end{tabular}} & $W=0$                               \\ \hline
\multicolumn{1}{|c|}{0.5} & \multicolumn{1}{c|}{1.5}    & \multicolumn{1}{c|}{-0.1} & \multicolumn{1}{c|}{12}                      & \multicolumn{1}{c|}{\begin{tabular}[c]{@{}c@{}}$w_1=+1$\\ $w_2=-1$\end{tabular}} & $W=0$                               \\ \hline
\multicolumn{1}{|c|}{0.5} & \multicolumn{1}{c|}{0.01}   & \multicolumn{1}{c|}{-0.1} & \multicolumn{1}{c|}{1}                       & \multicolumn{1}{c|}{\begin{tabular}[c]{@{}c@{}}$w_1=+1$\\ $w_2=-1$\end{tabular}} & $W=0$                               \\ \hline
\multicolumn{1}{|c|}{1}   & \multicolumn{1}{c|}{0.01}   & \multicolumn{1}{c|}{-0.1} & \multicolumn{1}{c|}{1}                       & \multicolumn{1}{c|}{\begin{tabular}[c]{@{}c@{}}$w_1=+1$\\ $w_2=-1$\end{tabular}} & $W=0$                               \\ \hline
\multicolumn{1}{|c|}{1.5} & \multicolumn{1}{c|}{0.01}   & \multicolumn{1}{c|}{-0.1} & \multicolumn{1}{c|}{0.5}                     & \multicolumn{1}{c|}{\begin{tabular}[c]{@{}c@{}}$w_1=+1$\\ $w_2=-1$\end{tabular}} & $W=0$                               \\ \hline
\end{tabular}
\caption{Winding numbers and topological numbers for higher order QED corrected EHAdS black hole in grand canonical ensemble for negative $a$}
\label{HOEHBH_Grand_Can_Neg_a}
\end{table}

\subsection{Positive $a$}
\begin{figure}[ht]
	\centering
		\begin{subfigure}{0.5\textwidth}
			\centering
			\includegraphics[width=0.9\linewidth]{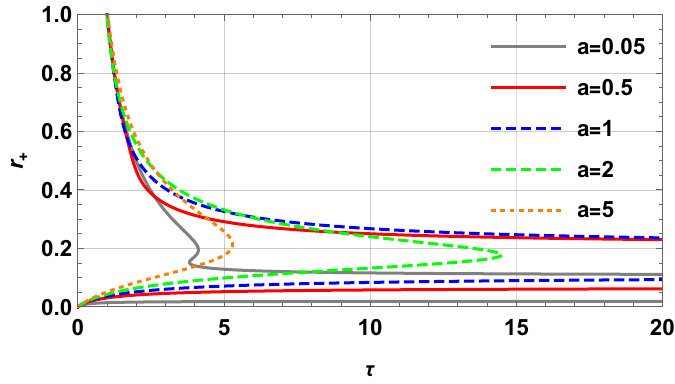}
			\caption{$\tau$ vs $r_{+}$ plot for different negative values of $a$. Here $P=0.5$ and $\phi_e=1$.}
			\label{Fig:HOEHBH_GrandCan_Positive_a_a_Variation}
		\end{subfigure}%
		\begin{subfigure}{0.5\textwidth}
			\centering
			\includegraphics[width=0.9\linewidth]{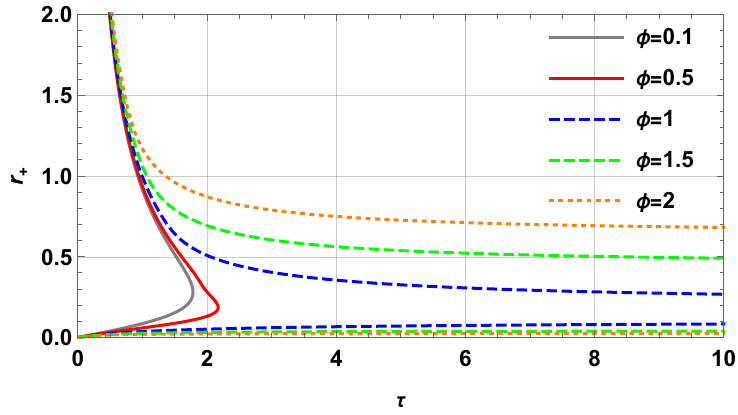}
			\caption{$\tau$ vs $r_{+}$ plot for different values of $\phi_e$. Here $a=1$ and $P=0.5$}
			\label{Fig:HOEHBH_GrandCan_Positive_a_phi_Variation}
		\end{subfigure} \\
		\begin{subfigure}{0.5\textwidth}
			\centering
			\includegraphics[width=0.9\linewidth]{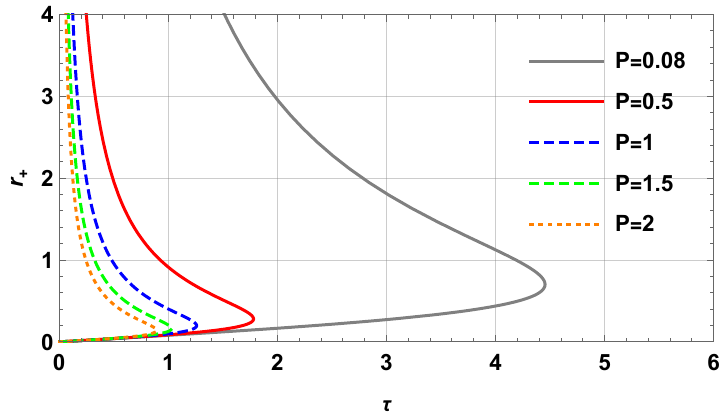}
			\caption{$\tau$ vs $r_{+}$ plot for different values of $P$. Here $a=1$ and $\phi_e=0.1$}
			\label{Fig:HOEHBH_GrandCan_Positive_a_P_Variation}
		\end{subfigure}%
	\caption{ $\tau$ vs $r_+$ plot with different parameter variations for higher order QED corrected EHAdS black hole in grand canonical ensemble (for positive $a$).}
	\label{Fig:HOEHBH_GrandCan_Tau_Plot_All_Positive}
	\end{figure}

\begin{figure}[h!]
	\centerline{
	\includegraphics[scale=0.6]{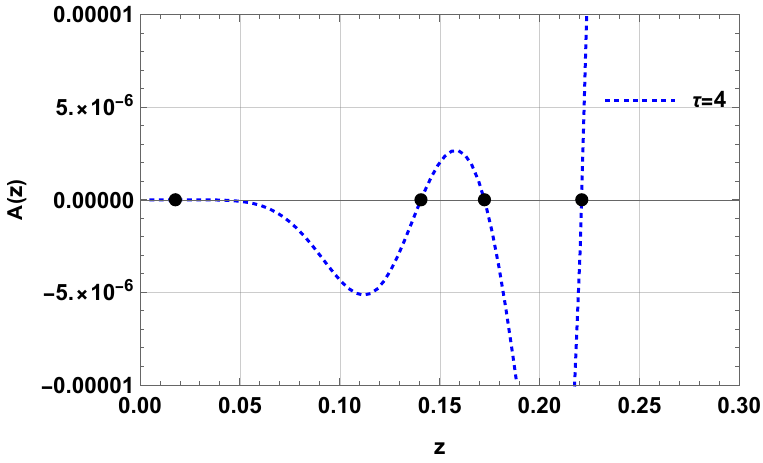}}
	\caption	{Plot of the polynomial function $\mathcal{A}(z)$ for $\tau=4$. The black dots represent the roots of the polynomial.}	\label{Fig:Poly_Plot_HOEHBH_GranCan_a_Pos_a}
	\end{figure}

For positive EH parameter $a$, the $\tau(r_+)$ curve is shown in \autoref{Fig:HOEHBH_GrandCan_Tau_Plot_All_Positive} which is observed to have either $2$ or $4$ different branches. In \autoref{Fig:HOEHBH_GrandCan_Positive_a_a_Variation} we observe four branches of the $\tau(r_+)$ curve for $(P,\phi_e,a,\beta)=(0.5,1,0.05,0.11)$. In this parameter combination the polynomial function is shown in \autoref{Fig:Poly_Plot_HOEHBH_GranCan_a_Pos_a}. For $\tau=4$ we have four roots of $\mathcal{A}(z)$ which are $z_1=0.221127$, $z_2=0.1724031$, $z_3=0.140630$ and $z_4=0.017681$. From residue method discussed earlier, the corresponding winding numbers are $w_1=+1$, $w_2=-1$, $w_3=+1$ and $w_4=-1$ yielding the topological number to be $W=+1-1+1-1=0$. For the other set of parameters, the $\tau$ curve has two branches with $w_1=+1$ and $w_2=-1$ winding number. The topological number is hence $W=+1-1=0$. The winding numbers and topological numbers for various set of parameters are shown in \autoref{HOEHBH_Grand_Can_Posi_a}.

\begin{table}[]
\begin{tabular}{|cccccc|}
\hline
\multicolumn{6}{|c|}{Higher order EHAdS black hole in grand canonical ensemble ($\beta = 0.11$)}                                                                                                                                                                                      \\ \hline
\multicolumn{3}{|c|}{Parameters}                                                     & \multicolumn{1}{c|}{\multirow{2}{*}{$\tau$}} & \multicolumn{1}{c|}{\multirow{2}{*}{Winding number}}                                                   & \multirow{2}{*}{Topological number} \\ \cline{1-3}
\multicolumn{1}{|c|}{P}    & \multicolumn{1}{c|}{$\phi$} & \multicolumn{1}{c|}{a}    & \multicolumn{1}{c|}{}                        & \multicolumn{1}{c|}{}                                                                                  &                                     \\ \hline
\multicolumn{1}{|c|}{0.5}  & \multicolumn{1}{c|}{1}      & \multicolumn{1}{c|}{0.05} & \multicolumn{1}{c|}{4}                       & \multicolumn{1}{c|}{\begin{tabular}[c]{@{}c@{}}$w_1=+1$\\ $w_2=-1$\\ $w_3=+1$\\ $w_4=-1$\end{tabular}} & $W=0$                               \\ \hline
\multicolumn{1}{|c|}{0.5}  & \multicolumn{1}{c|}{1}      & \multicolumn{1}{c|}{0.5}  & \multicolumn{1}{c|}{4}                       & \multicolumn{1}{c|}{\begin{tabular}[c]{@{}c@{}}$w_1=+1$\\ $w_2=-1$\end{tabular}}                       & $W=0$                               \\ \hline
\multicolumn{1}{|c|}{0.5}  & \multicolumn{1}{c|}{1}      & \multicolumn{1}{c|}{1}    & \multicolumn{1}{c|}{10}                      & \multicolumn{1}{c|}{\begin{tabular}[c]{@{}c@{}}$w_1=+1$\\ $w_2=-1$\end{tabular}}                       & $W=0$                               \\ \hline
\multicolumn{1}{|c|}{0.5}  & \multicolumn{1}{c|}{0.1}    & \multicolumn{1}{c|}{2}    & \multicolumn{1}{c|}{5}                       & \multicolumn{1}{c|}{\begin{tabular}[c]{@{}c@{}}$w_1=+1$\\ $w_2=-1$\end{tabular}}                       & $W=0$                               \\ \hline
\multicolumn{1}{|c|}{0.5}  & \multicolumn{1}{c|}{0.01}   & \multicolumn{1}{c|}{1}    & \multicolumn{1}{c|}{1}                       & \multicolumn{1}{c|}{\begin{tabular}[c]{@{}c@{}}$w_1=+1$\\ $w_2=-1$\end{tabular}}                       & $W=0$                               \\ \hline
\multicolumn{1}{|c|}{0.5}  & \multicolumn{1}{c|}{0.1}    & \multicolumn{1}{c|}{1}    & \multicolumn{1}{c|}{1}                       & \multicolumn{1}{c|}{\begin{tabular}[c]{@{}c@{}}$w_1=+1$\\ $w_2=-1$\end{tabular}}                       & $W=0$                               \\ \hline
\multicolumn{1}{|c|}{0.5}  & \multicolumn{1}{c|}{1.5}    & \multicolumn{1}{c|}{1}    & \multicolumn{1}{c|}{10}                      & \multicolumn{1}{c|}{\begin{tabular}[c]{@{}c@{}}$w_1=+1$\\ $w_2=-1$\end{tabular}}                       & $W=0$                               \\ \hline
\multicolumn{1}{|c|}{0.08} & \multicolumn{1}{c|}{0.65}   & \multicolumn{1}{c|}{1}    & \multicolumn{1}{c|}{5.3}                     & \multicolumn{1}{c|}{\begin{tabular}[c]{@{}c@{}}$w_1=+1$\\ $w_2=-1$\\ $w_3=+1$\\ $w_4=-1$\end{tabular}} & $W=0$                               \\ \hline
\multicolumn{1}{|c|}{0.5}  & \multicolumn{1}{c|}{0.1}    & \multicolumn{1}{c|}{1}    & \multicolumn{1}{c|}{1}                       & \multicolumn{1}{c|}{\begin{tabular}[c]{@{}c@{}}$w_1=+1$\\ $w_2=-1$\end{tabular}}                       & $W=0$                               \\ \hline
\multicolumn{1}{|c|}{1}    & \multicolumn{1}{c|}{0.1}    & \multicolumn{1}{c|}{1}    & \multicolumn{1}{c|}{1}                       & \multicolumn{1}{c|}{\begin{tabular}[c]{@{}c@{}}$w_1=+1$\\ $w_2=-1$\end{tabular}}                       & $W=0$                               \\ \hline
\end{tabular}
\caption{Winding numbers and topological numbers for higher order QED corrected EHAdS black hole in grand canonical ensemble for positive $a$}
\label{HOEHBH_Grand_Can_Posi_a}
\end{table}

\clearpage
\section{Conclusion}
\label{Conclusion}
We have studied the thermodynamic topology of 4D Euler-Heisenberg-AdS black hole in different ensembles without and with the higher order QED correction using generalized off-shell free energy. We calculate the winding number and the topological number for the aforementioned black hole systems for both negative and positive EH parameters $a$ in different ensembles. \\

For 4D EHAdS black hole in canonical ensemble with negative EH parameter $a$, the $\tau(r_+)$ curve, which represents the zero points of the vector field $\phi$, is observed to posses either $1$ or $3$ black hole branches depending on the parameters pressure $P$, charge $Q$ and EH parameter $a$. The total topological number is conserved ($W=+1$) regardless of the different values of the thermodynamic parameters. For EHAdS black hole in canonical ensemble with positive EH parameter $a$, the $\tau(r_+)$ curve is observed to have either $2$ or $4$ black hole branches depending on the thermodynamic parameters. The topological number in this case is $W=0$. Variation of the thermodynamic parameters do not have any impact on the topological number. This implies that the topological class of 4D EHAdS black hole in canonical ensemble changes depending on the signature of EH parameter $a$. We have tabulated our findings in \autoref{EHBH_Negative_a} and \autoref{EHBH_Positive_a}.\\

Next, we have studied the thermodynamic topology of the higher-order QED corrected Euler-Heisenberg-AdS black hole. For this black hole system, we set the parameter $\beta=0.11$ and study its thermodynamic topological properties for both the negative and positive values of EH parameter $a$. In canonical ensemble, for the negative $a$, the $\tau(r_+)$ curve is observed to have either $3$ or $1$ branches depending on the parameters $P$, $Q$ and $a$. The topological number is found to be $W=+1$. Changing the thermodynamic parameters do not have any impact on $W$ i.e., it is conserved.  For the positive $a$ case, we have observed either $3$ or $5$ different branches of the $\tau(r_+)$ curve. The total topological number is also $W=+1$ and it is found to be conserved. Therefore, we conclude that the higher-order QED corrected EHAdS black hole falls under the same topological class regardless of the nature of EH parameter $a$. The results are tabulated in \autoref{HOEHBH_Negative_a} and \autoref{HOEHBH_Positive_a}.\\

For EHAdS black hole in grand canonical ensemble, with both negative and positive EH parameter $a$, we have observed $2$ different branches of $\tau(r_+)$ curve, respectively corresponding to positive and negative winding number. The topological number is hence $W=0$. In case of the higher order QED corrected EHAdS black hole, for negative EH parameter $a$ and $\beta=0.11$, we have found $2$ different branches of the $\tau(r_+)$ curve. On the other hand, for the same, with positive EH parameter $a$ and $\beta=0.11$, we have found either $2$ or $4$ different branches of the $\tau(r_+)$ curve. The topological number for both the black hole systems in grand canonical ensemble is found to be $W=0$ which is independent of the parameters pressure $P$, electric potential $\phi_e$ and EH parameter $a$. This suggests that the EHAdS black hole and the higher order QED corrected EHAdS black hole in grand canonical ensemble belong to the same topological class. The winding numbers and topological numbers for various set of parameters are shown in \autoref{EHBH_Grand_Can_Neg_a}, \autoref{EHBH_Grand_Can_Posi_a}, \autoref{HOEHBH_Grand_Can_Neg_a} and \autoref{HOEHBH_Grand_Can_Posi_a}.\\

An interesting question whether the currently known thermodynamic topological classification will hold good for other black hole solutions in different theories of gravity. We plan to address this question in our future works.


\section{Declaration}
The preparation of this manuscript is not assisted with any funds.

\clearpage

\appendix*
\section{The complex functions and polynomial functions}
\subsection{The Euler-Heisenberg AdS black hole in grand canonical ensemble}
The complex function $\mathcal{R}(z)$ is given as
\begin{equation}
\label{EHBH_Grand_Can_ComF}
\mathcal{R}(z)=\frac{
\begin{aligned}
&	-10 A z^{11/3} \left(\sqrt[3]{15} z^{5/3} \phi  \sqrt[3]{9 \phi  a^2+A \sqrt{3}}+120 z^3\right) \sqrt[3]{9 \phi  a^2+A \sqrt{3}} \\
&	+486 \sqrt[6]{3} a^4 z \phi ^3 \left(\sqrt[3]{3} z^{5/3} \left(8 P \pi  z^2+1\right) \sqrt[3]{9 \phi  a^2+A \sqrt{3}}+14\times 5^{2/3} z^3 \phi \right) \\
&	-6 a^2 \Bigl\{ -27 A z^{8/3} \phi ^2 \sqrt[3]{9 \phi  a^2+A \sqrt{3}}-216 A P \pi  z^{14/3} \phi ^2 \sqrt[3]{9 \phi  a^2+A \sqrt{3}}+1200 z^{20/3} \phi  \sqrt{3} \sqrt[3]{9 \phi  a^2+A \sqrt{3}} \\
&	+55\times 3^{5/6} \sqrt[3]{5} z^{16/3} \phi ^2 \left(9 \phi  a^2+A \sqrt{3}\right)^{2/3}-400 \sqrt[6]{3} 5^{2/3} z^8-126\times 15^{2/3} A z^4 \phi ^3\Bigr\}) \\
&	+45 a^3 \Bigl\{ -16 \sqrt{3} z^{14/3} \left(8 P \pi  z^2+1\right) \phi  \sqrt[3]{9 \phi  a^2+A \sqrt{3}}+108 z^{14/3} \phi ^3 \sqrt{3} \sqrt[3]{9 \phi  a^2+A \sqrt{3}} \\
&	+9\times 3^{5/6} \sqrt[3]{5} z^{10/3} \phi ^4 \left(9 \phi  a^2+A \sqrt{3}\right)^{2/3}-260 \sqrt[6]{3} 5^{2/3} z^6 \phi ^2\Bigr\} \\
&	-5 a \Bigl\{ -324 A z^{14/3} \phi ^2 \sqrt[3]{9 \phi  a^2+A \sqrt{3}}+8 z^{14/3} \Bigl( 10\times 3^{5/6} \sqrt[3]{5} z^{8/3} \sqrt[3]{9 \phi  a^2+A \sqrt{3}} \\
&	+24 A P \pi  z^2+3 A\Bigr) \sqrt[3]{9 \phi  a^2+A \sqrt{3}}-27 \sqrt[3]{15} A z^{10/3} \phi ^3 \left(9 \phi  a^2+A \sqrt{3}\right)^{2/3}+148\times 15^{2/3} A z^6 \phi \Bigr\}
\end{aligned}}
{
\begin{aligned}
&	-10 A z^{11/3} \tau  \left(\sqrt[3]{15} z^{5/3} \phi  \sqrt[3]{9 \phi  a^2+A \sqrt{3}}+120 z^3\right) \sqrt[3]{9 \phi  a^2+A \sqrt{3}} \\
&	+45 a^3 \Bigl\{-16 \sqrt{3} z^{14/3} \Big( \tau +4 \pi  z (2 P z \tau -1)\Big) \phi  \sqrt[3]{9 \phi  a^2+A \sqrt{3}}+108 z^{14/3} \tau  \phi ^3 \sqrt{3} \sqrt[3]{9 \phi  a^2+A \sqrt{3}} \\
&	+9\times 3^{5/6} \sqrt[3]{5} z^{10/3} \tau  \phi ^4 \left(9 \phi  a^2+A \sqrt{3}\right)^{2/3} -260 \sqrt[6]{3} 5^{2/3} z^6 \tau  \phi ^2\Bigr\} \\
&	+486 \sqrt[6]{3} a^4 z \phi ^3 \Bigl\{ 4 \sqrt[3]{3} \pi  z^{8/3} (2 P z \tau -1) \sqrt[3]{9 \phi  a^2+A \sqrt{3}}+\tau  \left(\sqrt[3]{3} z^{5/3} \sqrt[3]{9 \phi  a^2+A \sqrt{3}}+14\times 5^{2/3} z^3 \phi \right)\Bigr\} \\
&	+6 a^2 \Bigl[-1200 \sqrt{3} z^{20/3} \tau  \phi  \sqrt[3]{9 \phi  a^2+A \sqrt{3}}+z^{8/3} \phi ^2 \Bigl\{-55 3^{5/6} \sqrt[3]{5} z^{8/3} \tau  \sqrt[3]{9 \phi  a^2+A \sqrt{3}} \\
&	+27 A \tau +108 A \pi  z (2 P z \tau -1)\Bigr\} \sqrt[3]{9 \phi  a^2+A \sqrt{3}}+126\times 15^{2/3} A z^4 \tau  \phi ^3+400 \sqrt[6]{3} 5^{2/3} z^8 \tau \Bigr] \\
&	-5 a \Bigl[96 A \pi  z^{17/3} (2 P z \tau -1) \sqrt[3]{9 \phi  a^2+A \sqrt{3}}+\tau  \Bigl\{24 A z^{14/3} \sqrt[3]{9 \phi  a^2+A \sqrt{3}}-324 A z^{14/3} \phi ^2 \sqrt[3]{9 \phi  a^2+A \sqrt{3}} \\
&	+80\times 3^{5/6} \sqrt[3]{5} z^{22/3} \left(9 \phi  a^2+A \sqrt{3}\right)^{2/3}-27 \sqrt[3]{15} A z^{10/3} \phi ^3 \left(9 \phi  a^2+A \sqrt{3}\right)^{2/3}+148\times 15^{2/3} A z^6 \phi \Bigr\}\Bigr]
\end{aligned}}
\end{equation}
 and the polynomial function $\mathcal{A}(z)$ is given as
 \begin{equation}
 \label{EHBH_Grand_Can_PolyF}
 \begin{aligned}
  \mathcal{A}(z)=
&	-10 A z^{11/3} \tau  \left(\sqrt[3]{15} z^{5/3} \phi  \sqrt[3]{9 \phi  a^2+A \sqrt{3}}+120 z^3\right) \sqrt[3]{9 \phi  a^2+A \sqrt{3}} \\
&	+45 a^3 \Bigl\{-16 \sqrt{3} z^{14/3} \Big( \tau +4 \pi  z (2 P z \tau -1)\Big) \phi  \sqrt[3]{9 \phi  a^2+A \sqrt{3}}+108 z^{14/3} \tau  \phi ^3 \sqrt{3} \sqrt[3]{9 \phi  a^2+A \sqrt{3}} \\
&	+9\times 3^{5/6} \sqrt[3]{5} z^{10/3} \tau  \phi ^4 \left(9 \phi  a^2+A \sqrt{3}\right)^{2/3} -260 \sqrt[6]{3} 5^{2/3} z^6 \tau  \phi ^2\Bigr\} \\
&	+486 \sqrt[6]{3} a^4 z \phi ^3 \Bigl\{ 4 \sqrt[3]{3} \pi  z^{8/3} (2 P z \tau -1) \sqrt[3]{9 \phi  a^2+A \sqrt{3}}+\tau  \left(\sqrt[3]{3} z^{5/3} \sqrt[3]{9 \phi  a^2+A \sqrt{3}}+14\times 5^{2/3} z^3 \phi \right)\Bigr\} \\
&	+6 a^2 \Bigl[-1200 \sqrt{3} z^{20/3} \tau  \phi  \sqrt[3]{9 \phi  a^2+A \sqrt{3}}+z^{8/3} \phi ^2 \Bigl\{-55 3^{5/6} \sqrt[3]{5} z^{8/3} \tau  \sqrt[3]{9 \phi  a^2+A \sqrt{3}} \\
&	+27 A \tau +108 A \pi  z (2 P z \tau -1)\Bigr\} \sqrt[3]{9 \phi  a^2+A \sqrt{3}}+126\ 15^{2/3} A z^4 \tau  \phi ^3+400 \sqrt[6]{3} 5^{2/3} z^8 \tau \Bigr] \\
&	-5 a \Bigl[96 A \pi  z^{17/3} (2 P z \tau -1) \sqrt[3]{9 \phi  a^2+A \sqrt{3}}+\tau  \Bigl\{24 A z^{14/3} \sqrt[3]{9 \phi  a^2+A \sqrt{3}}-324 A z^{14/3} \phi ^2 \sqrt[3]{9 \phi  a^2+A \sqrt{3}} \\
&	+80\times 3^{5/6} \sqrt[3]{5} z^{22/3} \left(9 \phi  a^2+A \sqrt{3}\right)^{2/3}-27 \sqrt[3]{15} A z^{10/3} \phi ^3 \left(9 \phi  a^2+A \sqrt{3}\right)^{2/3}+148\times 15^{2/3} A z^6 \phi \Bigr\}\Bigr]
\end{aligned}
 \end{equation}
where, $A=\sqrt{a^3 \left(27 a \phi ^2-40 z^2\right)}$.

\subsection{The higher order Euler-Heisenberg AdS black hole in grand canonical ensemble}
The complex function $\mathcal{R}(z)$ (as standard mathematica output) is given as
\begin{equation}
 \label{HOEHBH_Grand_Can_ComF}
\mathcal{R}(z)=
\frac{
\begin{aligned}
&	a \Bigl\{ 625 a \beta ^2+2 z^2 \left(2500 \pi  a \beta ^2 P+450 \beta -9\right)\Bigr\} \text{Root}\left[-\frac{5 \text{$\#$1}^5 a^2 \beta }{z^9}+\frac{6 \text{$\#$1}^3 a}{z^5}-\frac{60 \text{$\#$1}}{z}+60 \phi \&,1\right]^4 \\
&	-30 \Bigl[ 2 z^6 \big\{20 \beta  (3 \pi  a P+10)-3\bigr\} +15 a \beta  z^4\Bigr] \text{Root}\left[-\frac{5 \text{$\#$1}^5 a^2 \beta }{z^9}+\frac{6 \text{$\#$1}^3 a}{z^5}-\frac{60 \text{$\#$1}}{z}+60 \phi \&,1\right]^2 \\
&	+60 (550 \beta -3) z^7 \phi  \text{Root}\left[-\frac{5 \text{$\#$1}^5 a^2 \beta }{z^9}+\frac{6 \text{$\#$1}^3 a}{z^5}-\frac{60 \text{$\#$1}}{z}+60 \phi \&,1\right] \\
&	-1275 a \beta  z^3 \phi  \text{Root}\left[-\frac{5 \text{$\#$1}^5 a^2 \beta }{z^9}+\frac{6 \text{$\#$1}^3 a}{z^5}-\frac{60 \text{$\#$1}}{z}+60 \phi \&,1\right]^3+1500 \beta  z^8 \left(8 \pi  P z^2-15 \phi ^2+1\right)
\end{aligned}}
{\begin{aligned}
&	a \Bigl[ 625 a \beta ^2 \bigl\{4 \pi  z (2 P \tau  z-1)+\tau \bigr\} +18 (50 \beta -1) \tau  z^2\Bigr] \text{Root}\left[-\frac{5 \text{$\#$1}^5 a^2 \beta }{z^9}+\frac{6 \text{$\#$1}^3 a}{z^5}-\frac{60 \text{$\#$1}}{z}+60 \phi \&,1\right]^4 \\
&	-30 \Bigl[ 15 a \beta  z^4 \bigl\{ 4 \pi  z (2 P \tau  z-1)+\tau \bigr\} +2 (200 \beta -3) \tau  z^6\Bigr] \text{Root}\left[-\frac{5 \text{$\#$1}^5 a^2 \beta }{z^9}+\frac{6 \text{$\#$1}^3 a}{z^5}-\frac{60 \text{$\#$1}}{z}+60 \phi \&,1\right]^2 \\
&	+60 (550 \beta -3) \tau  z^7 \phi  \text{Root}\left[-\frac{5 \text{$\#$1}^5 a^2 \beta }{z^9}+\frac{6 \text{$\#$1}^3 a}{z^5}-\frac{60 \text{$\#$1}}{z}+60 \phi \&,1\right] \\
&	-1275 a \beta  \tau  z^3 \phi  \text{Root}\left[-\frac{5 \text{$\#$1}^5 a^2 \beta }{z^9}+\frac{6 \text{$\#$1}^3 a}{z^5}-\frac{60 \text{$\#$1}}{z}+60 \phi \&,1\right]^3+1500 \beta  z^8 \bigl\{ 4 \pi  z (2 P \tau  z-1)-15 \tau  \phi ^2+\tau \bigr\}
\end{aligned}}
\end{equation}
and the polynomial function $\mathcal{A}(z)$ (as standard mathematica output) is given as 
\begin{equation}
 \label{HOEHBH_Grand_Can_PolyF}
\begin{aligned}
\mathcal{A}(z)=
&	a \Bigl[ 625 a \beta ^2 \bigl\{4 \pi  z (2 P \tau  z-1)+\tau \bigr\} +18 (50 \beta -1) \tau  z^2\Bigr] \text{Root}\left[-\frac{5 \text{$\#$1}^5 a^2 \beta }{z^9}+\frac{6 \text{$\#$1}^3 a}{z^5}-\frac{60 \text{$\#$1}}{z}+60 \phi \&,1\right]^4 \\
&	-30 \Bigl[ 15 a \beta  z^4 \bigl\{ 4 \pi  z (2 P \tau  z-1)+\tau \bigr\} +2 (200 \beta -3) \tau  z^6\Bigr] \text{Root}\left[-\frac{5 \text{$\#$1}^5 a^2 \beta }{z^9}+\frac{6 \text{$\#$1}^3 a}{z^5}-\frac{60 \text{$\#$1}}{z}+60 \phi \&,1\right]^2 \\
&	+60 (550 \beta -3) \tau  z^7 \phi  \text{Root}\left[-\frac{5 \text{$\#$1}^5 a^2 \beta }{z^9}+\frac{6 \text{$\#$1}^3 a}{z^5}-\frac{60 \text{$\#$1}}{z}+60 \phi \&,1\right] \\
&	-1275 a \beta  \tau  z^3 \phi  \text{Root}\left[-\frac{5 \text{$\#$1}^5 a^2 \beta }{z^9}+\frac{6 \text{$\#$1}^3 a}{z^5}-\frac{60 \text{$\#$1}}{z}+60 \phi \&,1\right]^3+1500 \beta  z^8 \bigl\{ 4 \pi  z (2 P \tau  z-1)-15 \tau  \phi ^2+\tau \bigr\}
\end{aligned}
\end{equation}

\bibliographystyle{apsrev}
\end{document}